\documentclass[a4paper,11pt]{article}
\pdfoutput=1 

\usepackage[colorlinks=true, linkcolor = blue, urlcolor = blue, citecolor = blue, anchorcolor = blue]{hyperref}
\usepackage{amsmath,amssymb,mathtools,slashed}
\usepackage{multicol,multirow,makecell,caption}
\usepackage{graphicx,xcolor,float,easyReview}
\usepackage[normalem]{ulem}

\newcommand{\be}{\begin{equation}}
\newcommand{\ee}{\end{equation}}
\newcommand{\bea}{\begin{eqnarray}}
\newcommand{\eea}{\end{eqnarray}}
\newcommand{\cmmnt}[1]{\ignorespaces}

\newcommand{\bav}{\begin{array}{cccc}}

\makeatletter

\makeatletter
\newcommand*{\rom}[1]{\expandafter\@slowromancap\romannumeral #1@}
\makeatother

\begin{document}

\begin{flushright}HRI-RECAPP-2025-08\end{flushright}\medskip
 \begin{center} 
  {\Large\bf Exploring Fermionic Dark Matter in the Presence of Scalar Leptoquarks} \vskip .5cm
  {Shyamashish Dey\footnote{shyamashishdey@hri.res.in},
  Santosh Kumar Rai\footnote{skrai@hri.res.in},
\\[3mm]
  {\it{
   Regional Centre for Accelerator-based Particle Physics, Harish-Chandra Research Institute, \\
A CI of Homi Bhabha National Institute, 
Chhatnag Road, Jhunsi, Prayagraj 211019, India}\\
  }
}
 \end{center}
 \begin{abstract}
 We study an extension of the vector-like lepton dark matter model by introducing scalar leptoquarks that modify the properties of the vector-like lepton dark matter candidate, helping it evade stringent direct detection constraints. In the minimal setup, the neutral component of a pure $SU(2)$ doublet vector-like lepton fails to simultaneously account for the observed relic abundance while remaining consistent with current direct detection limits. We show that the addition of the scalar leptoquarks can induce corrections to the mass of the vector-like lepton dark matter, splitting it into two non-degenerate pseudo-Dirac states. This mass splitting can help evade the direct detection bounds naturally. In addition, the extended setup also opens up a larger parameter space of the model that can accommodate the correct relic density. 
\end{abstract}
   
\section{Introduction}\label{sec:intro}
The Standard Model (SM) of particle physics is the cornerstone of our understanding of subatomic phenomena, offering remarkably precise explanations. With the discovery of the Higgs boson at the Large Hadron Collider (LHC), all particles predicted by the SM have been experimentally confirmed. However, despite its many successes, the SM in its current form does not account for a viable dark matter (DM) candidate. Experimentally, DM has been confirmed to have a relic abundance of $\Omega h^2 = 0.12\pm0.001$ by observations made by Planck \cite{Planck:2018vyg}.  
In this paper, we will discuss a model of weakly interacting massive particle (WIMP) DM, as the name suggests, it has an interaction strength comparable to the electroweak interaction of the SM. Due to the DM-SM interaction, one can expect to observe the DM scattering with nucleons at present times in various direct detection (DD) experiments such as LUX-ZEPLIN~\cite{PRL.131.041002}, Pandax-4T \cite{PRL.127.261802}. There are also Indirect Detection (ID) facilities such as Fermi-LAT, MAGIC \cite{MAGIC_2016}, and H.E.S.S. \cite{PRL.129.111101} where the DM to SM conversion is being probed.

Due to the null observations of these detection facilities, there are severe bounds on the DM-SM interaction strength at different DM masses. Consequently, the primary challenge in building WIMP DM models is two-fold. First, DM must have sufficient interaction strength with SM particles to maintain thermal equilibrium in the early universe, enabling the freeze-out mechanism. Second, this interaction strength must be controlled by the model parameters so that the correct relic abundance is achieved while remaining consistent with the constraints from direct and indirect detection experiments. An interesting way to escape these constraints is to consider vector-like particles.  Because the Yukawa coupling of these particles with the SM Higgs can be chosen freely without being constrained by their mass, it allows us to control the Higgs-mediated interactions of DM with SM particles. This interaction contributes to the DM-nucleon scattering process in DD experiments. Vector-like particles appear in many extensions of the SM, such as grand unified theories (GUTs) like E6 \cite{Das:2017fjf}, $SO(10)$ \cite{Joglekar:2013zya}, from compactification in string theories \cite{Lebedev:2006kn}, as extensions of supersymmetric theories \cite{Martin:2009bg,Martin:2010dc,Martin:2012dg}, composite Higgs \cite{Chala:2018qdf}, Little Higgs~\cite{Arkani-Hamed:2002ikv} and in extra dimensions models \cite{Moreau:2006np,delAguila:2000kb}. There are also reviews on vector-like particles \cite{Ellis:2014dza,Alves:2023ufm}, which expand more on different possibilities from where these particles can arise.  Vector-like particles also have been used in various DM models where they can themselves provide a viable DM candidate \cite{Cohen:2011ec,Cirelli:2005uq,Dey:2022whc} or help achieve the correct DM relic density by introducing additional interactions in the model \cite{Bandyopadhyay:2023joz,Das:2024xle}.

One simple way to incorporate DM by introducing a vector-like particle is to consider a fermion doublet that transforms like any SM lepton doublet under the SM gauge symmetry. This vector-like lepton (VLL) doublet naturally provides a neutral fermion field that can serve as a viable candidate for DM. To stabilise the neutral component of the VLL, one needs to introduce a stabilising symmetry that prevents it from decaying to other particles. The most commonly used symmetry for this purpose is a discrete $Z_2$ group. Under this discrete $Z_2$ symmetry, SM particles are assigned positive parity (+1), while the VLLs receive negative parity (-1), ensuring that the lightest $Z_2$-odd particle remains stable and constitutes a viable dark matter candidate. It should be noted that, despite its simplicity, the pure VLL DM model faces severe phenomenological challenges to satisfy experimental observations of DM. 

In this work, we study the viability of the VLL component as DM by extending the minimal model with two scalar leptoquarks (LQ) to alleviate the shortcomings of the pure VLL DM model while retaining its essential features. The Scalar LQs are scalar fields which transform under the $SU(3)_C$ gauge group of the SM and couple leptons with quarks via Yukawa interaction. They appear in many well-motivated beyond standard model (BSM) scenarios such as GUT models \cite{Pati:1974yy,Georgi:1974sy,Fritzsch:1974nn}, R-parity violating SUSY \cite{Allanach:1999ic}, and Composite Higgs models \cite{Chala:2018qdf}. They have been utilised in describing many phenomena, notably for generating radiative neutrino mass \cite{Babu:2010vp}, Leptogenesis \cite{Fong:2013gaa}, and explaining the discrepancy in the anomalous magnetic moment of the muon \cite{Bauer:2015knc}. In this model, we impose the additional $Z_2$ symmetry on the LQ too, making them inert to direct couplings with SM fermions. They, however, help in generating mass splitting in the dark VLL sector, which helps satisfy DD bounds. One can also use the new DM-LQ interaction to open up new DM (co)annihilation processes to satisfy the DM relic density.

The paper is organised as follows. In Section \ref{sec:pureVLL}, we briefly review the pure $SU(2)$ doublet VLL model and discuss its shortcomings. In Section \ref{sec:model}, we describe the extension of the model and the relevant masses and mixing needed for our work. In Section \ref{sec:const}, we present the theoretical and experimental constraints on the model parameters, followed by the radiative corrections to the VLL mass and the DM phenomenology in Sections \ref{sec:vlmsplit} and \ref{sec:dmpheno}, respectively. We finally summarise our findings and conclude in Section \ref{sec:concl}.

\section{Status of pure VLL dark matter}\label{sec:pureVLL}	
In this section, we discuss the minimal scenario in which the DM candidate is a pure VLL, without including any additional new states. We will try to highlight why a purely VLL DM candidate in the minimal scenario faces challenges in achieving the observed relic abundance while remaining consistent with stringent limits from direct and indirect detection experiments.
In the minimal extension, we extend the Standard Model by a single vector-like lepton $f=f_L +f_R$, where $f_{L/R}$ are its chiral components, transforming as an $SU(2)_L$ doublet and with hypercharge $-1$. Similar to the SM lepton doublets, this yields one charged and one neutral component. The neutral component becomes the DM candidate by imposing an unbroken $Z_2$ symmetry under which the VLL is odd while all the SM fields are even  (Table~\ref{tab:vll}), forbidding its mixing with SM leptons and preventing it from decaying.

\begin{table}[H]
	\centering
		\begin{tabular}{|>{\centering\arraybackslash} m{3.0cm}|>{\centering\arraybackslash}m{1.25cm}>{\centering\arraybackslash}m{1.25cm}>{\centering\arraybackslash}m{1.25cm}c|}\hline
			
			\textbf{Fields}&  \multicolumn{3}{c}{\boldmath $\mathbf{\underbrace{ SU(3)_C\otimes SU(2)_L \otimes U(1)_{Y}}}$} & {\boldmath $\otimes\;Z_2$} \\ \hline\hline
			\multicolumn{5}{|l|}{Fermions:} \\ \hline
			$f_L=\left(\begin{matrix} f^0 \\ f^-  \end{matrix}\right)_L$ & 1  & 2 & $-$1 & $-$ \\
			\hline
			$f_R=\left(\begin{matrix} f^0 \\ f^-  \end{matrix}\right)_R$ & 1  & 2 & $-$1 & $-$ \\
			\hline
        \end{tabular}
		\caption{\it Here $f=f_L + f_R$ is the VLL. The definition of electromagnetic charge for all the fields follows the equation $Q_{EM} = T_3 + \frac{Y}{2}$. }
		\label{tab:vll}
	\end{table}
The VLL interacts solely with the SM gauge sector, and owing to its vector-like nature, a Dirac mass term is allowed at tree level. The relevant Lagrangian can be written as
\begin{equation}
 \mathcal{L}_{f} = i \, \bar{f}\gamma^\mu D_\mu \, f - m_f~\bar{f}\,f    
\end{equation}
where $D_\mu$ is the covariant derivative for an $SU(2)$ doublet with hypercharge $-1$. As both the neutral and charged components of the VLL do not mix with SM fermions, they are degenerate in mass. However, this degeneracy is lifted through radiative corrections from gauge interactions that induce an $\mathcal{O}$(MeV) mass splitting between the charged and neutral components of the VLL. This arises from one-loop diagrams mediated by $W$, $Z$, and photon exchange~\cite{Thomas:1998wy}, and can be expressed as
\begin{equation} \label{fig:vllpm}
    \delta m_{\pm}=\frac{\alpha}{2}m_Z f(m^2_f/m^2_Z)  
\end{equation}
where the function $f(r)$ is given by
\begin{equation}
   f(r) =\frac{\sqrt{r}}{\pi} \int_{0}^{1}dx \, (2 - x) \, ln\left(1 + \frac{x}{r(1 - x)^2}  \right) \,\,\, .
\end{equation}

\begin{figure}[H]
		\centering
		\includegraphics[angle=0,width=0.47\textwidth]{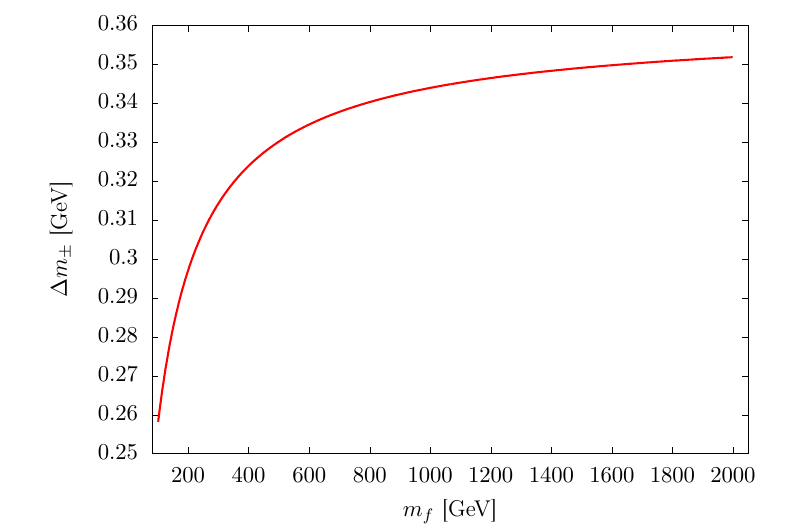} 
		
		\caption{Variation of mass gap between charged and neutral component of the VLL with tree-level mass parameter $m_f$. }
		\label{fig:vllmass}
	\end{figure}
The variation in the mass gap as a function of its tree-level mass parameter $m_f$ is shown in Fig.~\ref{fig:vllmass}. We note that for $m_f\gg m_Z$, $\delta m_{\pm}$ approaches its asymptotic value of $\simeq 355$ MeV. As a 
consequence, the charged VLL component remains heavier than the neutral one, ensuring the lighter neutral component acts as the DM candidate.

\begin{figure}[h!]
	\centering
	\includegraphics[angle=0,width=0.47\textwidth]{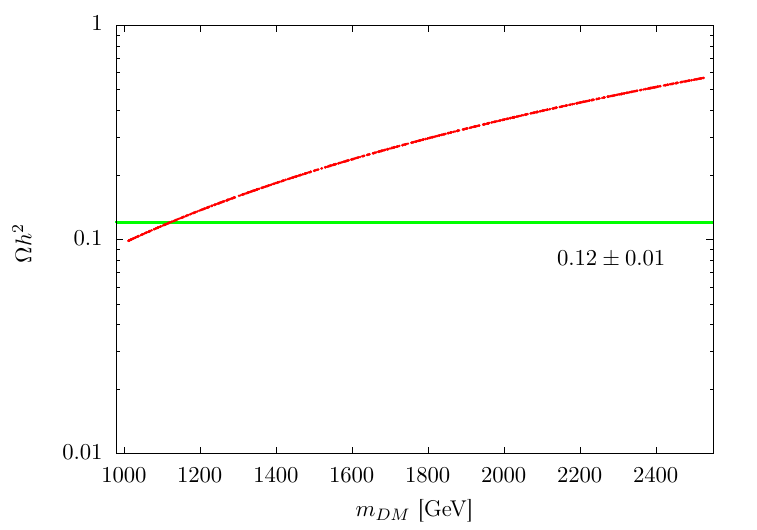} 
	\includegraphics[angle=0,width=0.48\textwidth]{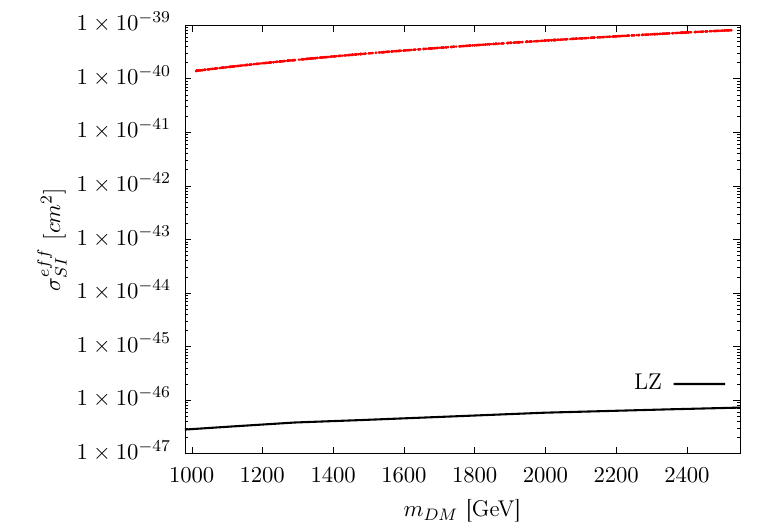}\\
	(A) \hspace{7.5cm} (B)
	\caption{Plot of DM relic density (A) and spin-independent direct detection cross-section (B) with varying VLL DM mass is given by the red curve. While the green region represents the DM relic density observed by Planck \cite{Planck:2018vyg} and the black curve is the DD bound given by LUX-ZEPLIN \cite{LZ:2024zvo}. Here, the direct detection cross-section has been scaled with the VLL relic density divided by the observed DM relic density.}
	\label{fig:vllrelic}
	\end{figure}

We now turn to the relic density and direct detection prospects of a pure VLL dark matter scenario. Since the VLL couples only to the SM gauge bosons, all relevant (co)annihilation channels,
\begin{equation}
\bar{f}^{0/\pm} f^{0/\mp} \to V \to {\rm SM \, SM}; \,\,\,\, V \in \{ W^{\pm}, Z, \gamma \} \,\,\,,
\end{equation}
are governed entirely by the fixed electroweak couplings. Consequently, the only free parameter in the dark sector (DS) is the mass spectrum of the VLL states. This constraint implies that the observed relic density can be reproduced only for a finely tuned DM mass of $\simeq 1.2$ TeV, as shown in 
Fig.~\ref{fig:vllrelic}(A). For $m_f < 1.2$ TeV, the DM is always underabundant when compared to the Planck \cite{Planck:2018vyg} measurement (green line in Fig.~\ref{fig:vllrelic}(A)), while for $m_f > 1.2$ TeV, it becomes overabundant, violating the relic density bound.

Being a Dirac fermion, the VLL also couples to the $Z$ boson, resulting in sizable contributions to direct detection via 
$Z$-exchange processes with nucleons. The resulting 
spin-independent scattering cross section significantly exceeds the current bounds from LUX-ZEPLIN~\cite{PRL.131.041002}, as shown in Fig.~\ref{fig:vllrelic}(B), thereby excluding the pure VLL DM scenario. 
Thus, although the pure VLL DM model is minimal and naturally provides a stable DM candidate, its viable parameter space is extremely restricted by relic density requirements and is entirely excluded by current direct detection limits.    
%

\section{VLL and inert LQ extension of SM}\label{sec:model}
As discussed in Sec.\ref{sec:pureVLL}, the minimal pure VLL model fails to simultaneously satisfy the observed relic density and direct detection constraints mainly because of its Dirac nature and $Z$ mediated interaction. A well-known mechanism to evade direct detection bounds involves considering a pseudo-Dirac fermion as the dark matter candidate, where the two Weyl components acquire a small Majorana mass splitting. If this mass gap exceeds the kinematic threshold for inelastic scattering at direct detection experiments, the DM escapes DD observation\cite{Tucker-Smith:2001myb}. For example, in our scenario, the inelastic scattering is dependent on the $Z$ boson mediated processes. Several extensions \cite{Bhattacharya:2015qpa,Arina:2012aj,Frank:2025jjt} have been proposed where this mechanism is employed to satisfy the severe DD bound of VLL DM. A typical approach introduces an $SU(2)_L$ triplet scalar with a small vacuum expectation value (vev) $v_t$ generated during electroweak symmetry breaking. Through additional Yukawa couplings, the VLL can interact with this triplet, opening up a wider DM mass range consistent with relic density observations. In addition, the neutral VLL acquires a small Majorana mass term proportional to $v_t$, making it a pseudo-Dirac fermion. However, the triplet scalar vev is strongly constrained by electroweak precision observables to small values. 
In this work, we propose an alternative framework in which an inert scalar leptoquark (LQ)  plays the key role in achieving the correct DM relic density while satisfying both DD and ID limits.

\subsection{Model} 
We extend the pure VLL model with two inert LQ scalars $S$ and $R$ with the charge assignment given in Table \ref{tab:model}. Here $S$ is an $SU(2)$ singlet while $R$ is a doublet, and both are odd under the $Z_2$ discrete symmetry which stabilises the DM component of the VLL $f$. 
\begin{table}[H]
	\centering
		\begin{tabular}{|>{\centering\arraybackslash} m{3.0cm}|>{\centering\arraybackslash}m{1.25cm}>{\centering\arraybackslash}m{1.25cm}>{\centering\arraybackslash}m{1.25cm}c|}\hline
			
			\textbf{Fields}&  \multicolumn{3}{c}{\boldmath $\mathbf{\underbrace{ SU(3)_C\otimes SU(2)_L \otimes U(1)_{Y}}}$} & {\boldmath $\otimes\;Z_2$} \\ \hline\hline
			\multicolumn{5}{|l|}{Fermions:} \\ \hline
			$Q_L=\left(\begin{matrix} u \\ d \end{matrix}\right)_{L}$ & 3 & 2 & $\phantom{-}\frac{1}{3}$ & $+$\\
			\hline
			$u_R$ & 3 & 1 & $\hphantom{-}\frac{4}{3}$ & $+$ \\[1pt]
			\hline
			$d_R$ & 3  & 1 & $-\frac{2}{3}$ & $+$ \\
			\hline
			$L_L=\left(\begin{matrix} \nu_\ell \\ \ell  \end{matrix}\right)_L$ & 1  & 2 & $-$1 & $+$ \\
			\hline
			$L_R$ & 1  & 1 & $-2$ & $+$ \\
			\hline
			$f_L=\left(\begin{matrix} f^0 \\ f^-  \end{matrix}\right)_L$ & 1  & 2 & $-$1 & $-$ \\
			\hline
			$f_R=\left(\begin{matrix} f^0 \\ f^-  \end{matrix}\right)_R$ & 1  & 2 & $-$1 & $-$ \\
			\hline\hline 
			\multicolumn{5}{|l|}{Scalars:} \\ \hline
			$ H = \left(\begin{matrix} H^+ \\ H^0 \end{matrix}\right)$ &  1 & 2 & \phantom{-}1 & $+$ \\
			\hline
			$R = \left(\begin{matrix} R_u \\ R_d \end{matrix}\right)$ &  3 & 2 & \phantom{-} $\frac{1}{3}$ & $-$ \\
			\hline
			$S $ &  3 & 1 & \phantom{-}-$\frac{2}{3}$ & $-$ \\
			\hline
		\end{tabular}
		\caption{\it The fields of the model considered and their quantum number under the SM gauge group and the additional $Z_2$ symmetry. Here $f=f_L + f_R$ is the VLL, and $R$ and $S$ are the $SU(2)_L$ doublet and singlet scalar LQs.}
		\label{tab:model}
	\end{table}

 We will see in the following sections that the extra Yukawa interactions of the VLL with the LQs provide additional annihilation processes for the VLL and allowing it to satisfy the relic density bounds for a large region of DM mass. We will also show that LQ will provide Majorana mass to the neutral VLL component via one-loop diagrams, which splits it into two pseudo-Dirac fermions, and hence the DD bounds can be satisfied.
\subsection{Lagrangian}
The scalar potential involving the leptoquark and the Standard Model Higgs doublet is given by 
\begin{align}\label{eq:pot}
	V(H,R,S) = -\mu_1^2 {H}^\dag H  + \mu_R^2 {R}^\dag R + + \mu_S^2 {S}^\dag S + \lambda_1 ({H}^\dag H)^2 
	+ \lambda_2 ({R}^\dag R)^2 + \lambda_3( {S}^\dag S)^2 \nonumber \\  + \lambda_4 ({R}^\dag R)({S}^\dag S)+ \lambda_5( {H}^\dag H)({R}^\dag R) + \lambda_6({R}^\dag H)( {H}^\dag R)+\lambda_7 ({H}^\dag H)({S}^\dag S) \nonumber \\ +\mu_{RS}\left[ {R}^\dag H S+ h.c.\right] \,\,\,. 	
\end{align} 
This potential contains a cubic interaction term between the LQs and the SM Higgs doublet. Following electroweak symmetry breaking, when the SM Higgs acquires its vacuum expectation value (vev), this interaction induces mixing among the down-type LQs. The resulting mixing alters the mass spectrum of the LQ states and affects their decay channels, which results in specific collider signatures and phenomenological implications of the model.
 
The Yukawa interaction between the VLL and LQs and the VLL tree-level mass term is given by
\begin{equation}\label{eq:yukawa}
	\mathcal{L}_{f} = y_{a} ~ \overline{Q_L^c} f_L S^\dagger + y_{b} ~ \overline{f_L} R^\dagger d_R + y_{c} ~ \overline{u_R^c} R^\dagger f_R - m_f~\bar{f} \, f + h.c. \,\,\,.
\end{equation}	
It is worth noting from Eq.~\ref{eq:yukawa} that the Yukawa interactions proportional to $y_a$ and $y_c$ violate fermion number by two units, whereas the term proportional to $y_b$ preserves fermion number. The presence of fermion-number–violating couplings can have important phenomenological implications, such as contributing to processes that induce baryon or lepton number violation, while the fermion-number–conserving interaction helps understand the standard decay channels and mixing patterns. 
%
\subsection{LQ mass and mixing angle}
To quantify the mixing among the leptoquarks, we construct the mass matrix for the down-type LQs and their physical eigenstates, after diagonalization. These mass eigenstates, and the associated mixing patterns, will play a crucial role in determining both low-energy constraints and the direct search limits from the LHC, which we analyze in subsequent sections. 
The tree-level mass matrix of the down-type LQs is shown in Eq.~\ref{eq:scalarmassmatrix}, where the basis is chosen as $\Sigma=\left( S \, \, R_d \right)$. 
		\begin{equation}\label{eq:scalarmassmatrix}
			M_{\Sigma^d}^2=\left(\begin{matrix} \mu_S^2+\frac{\lambda_6 v^2}{2} ~~~~~~~~\frac{\mu_{RS} v}{\sqrt{2}}  \\ \frac{\mu_{RS} v}{\sqrt{2}}~~~~\mu_R^2+\frac{(\lambda_4+\lambda_5) v^2}{2} \end{matrix}\right) \,\,\,.
		\end{equation}
Here $v$ is the vev of the SM Higgs doublet $H$. We then diagonalise the mass matrix by redefining the fields as $\Sigma=U_{RS}\Sigma^d$, where $U_{RS}$ is a unitary matrix given by
		\begin{equation}\label{eq:scalarmixing}
			U_{RS}=\left(\begin{matrix} \cos\theta ~~~~\sin\theta\\ -\sin\theta~~~~\cos\theta \end{matrix}\right) \,\,\,.
		\end{equation}

The mass eigenvalue of the down-type LQ is
	\begin{equation}\label{eq:scalarmass}
		m_{\Sigma_{1(2)}^{d}}^2=\frac{1}{4}\bigg[2\mu_S^2+2\mu_R^2 +(\lambda_4+\lambda_5+\lambda_6)v^2 \pm \sqrt{(2\mu_R^2-2\mu_S^2 +(\lambda_4+\lambda_5-\lambda_6)v^2)^2+8v^2\mu_{RS}^2}\bigg]\,. 
	\end{equation}

We can also express the angle $\theta$ in terms of the diagonalized LQ masses as
	\begin{equation}\label{eq:scalarangle}
	   \sin 2\theta=\frac{ \sqrt{2}\mu_{RS} v}{m_{\Sigma_{2}^{d}}^2-	m_{\Sigma_1^{d}}^2} \,\,\,.
	\end{equation}
We rename the up type LQ $R^u$ to $\Sigma^u$, and it's mass is given by $m_{\Sigma^{u}}^2=\mu_R^2+\frac{\lambda_4 v^2}{2}$. In the expression of LQ masses we considered the couplings $\lambda_4$, $\lambda_5$ and $\lambda_6$ negligibly small as they do not play a significant role in DM phenomenology.
Thus the LQ masses can be written as
\begin{equation}
    m_{\Sigma_{1(2)}^{d}}^2\simeq\frac{1}{2}\bigg[\mu_S^2+\mu_R^2  \pm \sqrt{(\mu_R^2-\mu_S^2 )^2+2v^2\mu_{RS}^2}\bigg], 
    \qquad  m_{\Sigma^u}^2\simeq\mu^2_R \,\,\, .
\end{equation}
	\section{Theoretical and Experimental Constraints}\label{sec:const}
  Before calculating the corrections to the $Z_2$ odd VLL mass eigenstates at one-loop, we first outline the regions of parameter space of the model that are consistent with different theoretical and experimental bounds, and relevant for the loop contribution.
\paragraph{Vacuum Stability --} The tree-level vacuum stability conditions are similar to those of a $Z_3$ DM model, which has been calculated in Ref.~\cite{Kannike:2016fmd}.  
	Thus, the positivity requirement of the ground state leads to the following conditions given below.
	\begin{align}\label{eqn:BFB}
	    & \lambda_{1,2,3}>0, 
         \quad \overline{\lambda_5} \equiv \lambda_5 + \sqrt{\lambda_1 +\lambda_2}>0 \nonumber \\ 
        & \overline{\lambda_7} \equiv \lambda_7 + \sqrt{\lambda_1 +\lambda_3}>0, \quad \overline{\lambda_4} \equiv \lambda_4 + \sqrt{\lambda_1 +\lambda_2}>0\\ 
        &\lambda_5\sqrt{\lambda_3}+\lambda_7\sqrt{\lambda_1}+
        \lambda_4\sqrt{\lambda_2}+\sqrt{\lambda_1 \lambda_2 \lambda_3}+\sqrt{\overline{\lambda_4}\,\overline{\lambda_5} \,\overline{\lambda_7}}>0 \,\,\, .\nonumber
	\end{align}    
	We note that all the conditions in Eq. \ref{eqn:BFB} can be simultaneously satisfied by choosing non-negative values for the quartic couplings. However, all of these couplings cannot be taken arbitrarily small. As emphasized in Ref.~\cite{Babu:2010vp}, the one-loop corrections to the quartic couplings given by Eq.~\ref{eqn:quartic}, can be negative. 
    \begin{align} \label{eqn:quartic}
        \lambda^{1L}_{1} = 12\, \mathcal{\zeta}(\mu_R,\mu_S,\mu_{RS}), &&
        \lambda^{1L}_{2} =  \mathcal{\zeta}(\mu_R,m_h,\mu_{RS}), &&
        \lambda^{1L}_{3} =  \mathcal{\zeta}(\mu_S,m_h,\mu_{RS}) \,.
    \end{align}
    Here $\mu_R,\mu_S$ and $\mu_{RS}$ are the mass parameters of the scalar potential \ref{eq:pot} and function $\mathcal{\zeta}$ is given by 
    \begin{equation}
        \mathcal{\zeta}(x,y,z)=-\frac{1}{128\pi^2}\frac{z^4}{x^2-y^2}\left[ \frac{x^2+y^2}{x^2-y^2}ln\left(\frac{x^2}{y^2}\right)-2 \right] \,\,\,.
    \end{equation}    
    If the magnitude of these corrections exceeds the corresponding tree-level values, the scalar potential becomes unstable, thereby spoiling vacuum stability. To ensure stability in our setup, we restrict to the LQ mass range $1000~\text{GeV} < m_\Sigma < 3000~\text{GeV}$ and take $\lambda_{2,3} = 0.01$. This choice is sufficiently larger than the expected negative one-loop contributions and guarantees the stability of the potential over the parameter space relevant for our analysis 

\paragraph{Perturbativity --} 
In order to ensure the reliability of perturbative calculations, we require that all coupling parameters remain within the perturbative regime. This guarantees that higher-order corrections to physical observables are progressively suppressed. Thus, the scalar quartic couplings are constrained by the standard perturbativity bound,  
\begin{equation}
|\lambda_i| \lesssim 4\pi, \quad i = 1,2,3,4,5,6,7 \,\,\,,
\end{equation}
which follows from the requirement that tree-level unitarity is not violated.  
Similarly, the Yukawa couplings are also required to satisfy the perturbativity condition,  
\begin{equation}
|y_{a,b,c}| \lesssim \sqrt{4\pi}\,\,\,.
\end{equation}

%


\paragraph{Electroweak precision observables --}  
The parameter space of the model is further constrained by electroweak precision observables (EWPOs), in particular the oblique parameter $T$. Here, we show the contributions to the $T$ parameter considering only the contributions from the BSM fields. The current global fit gives \cite{ParticleDataGroup:2024cfk}
\begin{equation}
T = 0.01 \pm 0.12 \,\,\,. 
\end{equation}  
Hence, the additional contribution from the LQ must satisfy  
\begin{equation}
|\Delta T_{\text{LQ}}| \lesssim 0.12 \,\,\,,
\end{equation}
to remain compatible with experimental bound.  

For scalar multiplets, the contribution to $\Delta T$ is sensitive to mass splittings among the components of the $SU(2)_L$ multiplet. The LQ contribution to the $T$ parameter is therefore given by
\begin{equation}
   \Delta T = \frac{3g_2^2}{32\pi^2 m^2_W}\left[ \cos^2(\theta)\mathcal{F}(m^2_{\Sigma^u},m^2_{\Sigma^d_1})+\sin^2(\theta)\mathcal{F}(m^2_{\Sigma^u},m^2_{\Sigma^d_2})-\frac{\sin^2(2\theta)}{4}\mathcal{F}(m^2_{\Sigma^d_1},m^2_{\Sigma^d_2})\right]
\end{equation}
where the function $\mathcal{F}(x,y)$ is given by
\begin{equation}
	\mathcal{F}\left(x, y\right)= \begin{cases} \frac{x+y}{2} - \frac{xy}{x-y} \ln\left(\frac{x}{y}\right), & \text{if } x\neq y,\\
		0,              & \text{if } x = y.\end{cases}
\end{equation}
Finally, we note that the VLLs are assumed to have degenerate charged and neutral components at the tree level. As a result, they do not  contribute to the $T$ parameter.
	\begin{figure}[H]
		\centering
		\includegraphics[angle=0,width=0.58\textwidth]{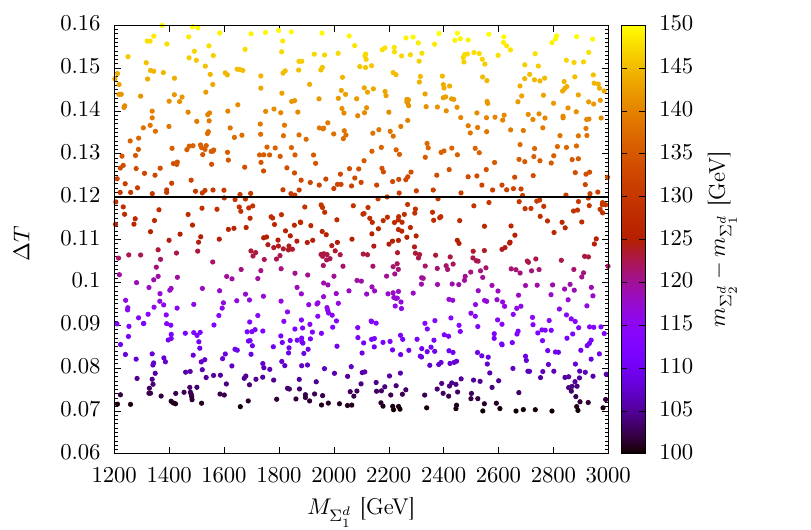} 
		\caption{ Variation of $\Delta T$ with the scalar LQ mass spectrum, assuming maximal mixing ($\sin 2\theta = 1$) between the down-type states. The black line denotes the experimental upper bound on $\Delta T$.}
		\label{fig:ewpt}
	\end{figure}
From the colour bar in Fig.~\ref{fig:ewpt}, we observe that the electroweak precision bound from the $T$ parameter allows a maximum mass splitting of approximately $\Delta M\simeq 130~\text{GeV}$ between the two down-type LQ states when maximal mixing is considered ($\theta = \pi/4$). For smaller values of $\sin 2\theta$, the constraint from the $T$ parameter becomes less stringent, thereby permitting a larger mass splitting in those scenarios. However, in our analysis, we adopt a conservative range by assuming a fixed benchmark splitting of  
\begin{equation}
M_{\Sigma^d_2} - M_{\Sigma^d_1} = 128~\text{GeV} \,\,\,,
\end{equation}
independent of the value of $\sin 2\theta$, to remain well within the allowed region of parameter space.

\paragraph{Direct production bound --} Since the LQs in our model carry a $Z_2$ charge, they do not couple to SM leptons nor behave as diquarks. Therefore, the conventional bounds on ordinary LQs, those without such a discrete symmetry do not directly apply here. However, the phenomenology of our LQs closely resembles that of squarks in supersymmetric scenarios, where squarks decay into a neutralino and an SM quark. In our case, the analogous decay is that of an LQ into a DM particle and an SM quark. Motivated by this similarity, we adopt the current experimental limits \cite{ATLAS:2024lda} on squark masses to constrain the mass of the LQs in our model as given below

\begin{equation}
    m_{\Sigma^u}> 600\, \mathrm{GeV}, \quad m_{\Sigma^d}> 650\, \mathrm{GeV} \,\,\,.
\end{equation}
\paragraph{Higgs signal bound --}  
As our model includes charged LQ, they can lead to modifications of the Higgs diphoton decay rate, $h \rightarrow \gamma\gamma$, through additional one-loop triangle diagrams involving LQs. This decay mode is a sensitive probe of new physics, as the Higgs signal strength $\mu$ is precisely measured \cite{ATLAS:2022vkf}, and it is given by
\begin{equation}
\mu = \dfrac{\sigma^{\mathrm{BSM}}_h\times BR^{\mathrm{BSM}}_{h\rightarrow\gamma\gamma}}{\sigma^{\mathrm{SM}}_h\times BR^{\mathrm{SM}}_{h\rightarrow\gamma\gamma}} = 1.11 \pm 0.09 \,\,\, .
\end{equation}
To compute the allowed mass and the interaction strength of the LQs to the SM Higgs, we evaluate the modified $h \rightarrow \gamma\gamma$ branching ratio in our model as a function of the down-type LQ mixing angle and the corresponding mass splitting, with the dependence controlled by the cubic scalar interaction $\mu_{RS}$, which couples the SM Higgs doublet to the LQs (see Eq.~\ref{eq:scalarangle}). This provides direct collider constraints on the parameter space.  

To calculate the Higgs to diphoton decay width in the presence of LQs we follow the methodology given in Ref. \cite{Djouadi:1998az}, which is given by  
\begin{equation}\label{eqn:haawidth}
    \Gamma(h\rightarrow\gamma\gamma) = \dfrac{\alpha^2 m_h^3}{256 \pi^3v^2} \left|F_1(\tau_W)+\sum_f N^C_{f}Q_f^2 F_{1/2}(\tau_f)+\sum_\Sigma N^C_{\Sigma}Q_\Sigma^2\,g_{\Sigma}F_0(\tau_\Sigma)\right|^2 \,\,\, .
\end{equation}
Here $\tau$ for all particles ($W^\pm$, $q$, $\ell^\pm$ and $\Sigma$) in the loop is defined as 
\begin{equation}
    \tau_i = \dfrac{4M_i^2}{M_h^2} \,\,\, .
\end{equation}
Now the loop factor are given by
\begin{align}
    F_1=&2+3\tau+3\tau(2-\tau)f(\tau),\\
    F_{1/2}=&-2\tau\left[1+(1-\tau)f(\tau)\right],\\
    F_0=&\tau\left[1-\tau f(\tau)\right]\,\,\,  .
\end{align}
While the function $f(\tau)$ is given by
\begin{equation}
    f(\tau)= \left\{\begin{matrix} \left[\sin^{-1}\left(\dfrac{1}{\sqrt{\tau}}\right)\right]^2, \hspace{2.4cm}\tau\geq1\\
    -\dfrac{1}{4}\left[\ln{\left(\dfrac{1+\sqrt{1-\tau}}{1-\sqrt{1-\tau}}\right)} -i\pi\right]^2,\,\, \tau< 1
    \end{matrix}\right.
\end{equation}

The LQ-Higgs coupling in Eqn. \ref{eqn:haawidth} is given by 
\begin{equation}
   g_{\Sigma^d_{1,2}}=\mp\dfrac{1}{\sqrt{2}}\mu_{RS}\sin{2\theta} \,\,\, .
\end{equation}
In the above equation we are taking the value of $\lambda_4, \lambda_5$ and $\lambda_6$ negligibly small as it further increases the $h\rightarrow\gamma\gamma$ decay width because of the extra loop contributions from up-type LQ, $\Sigma^u$.

	\begin{figure}[H]
		\centering
		\includegraphics[angle=0,width=0.45\textwidth]{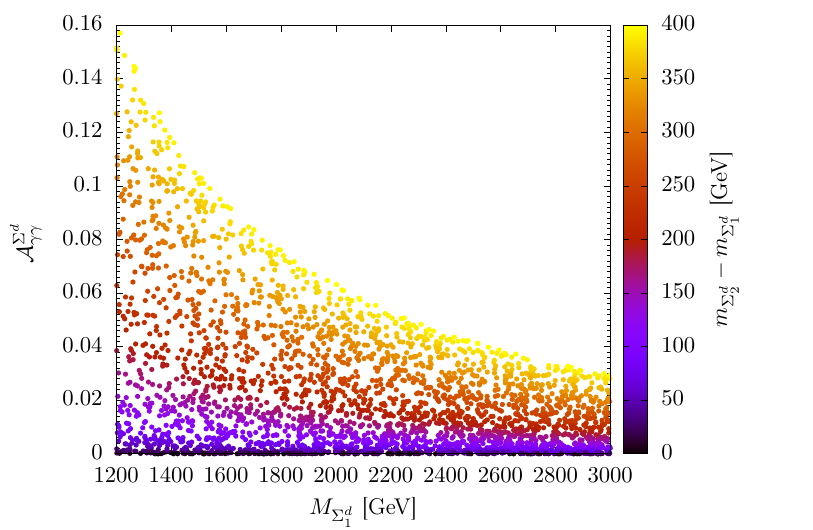} 
        \includegraphics[angle=0,width=0.45\textwidth]{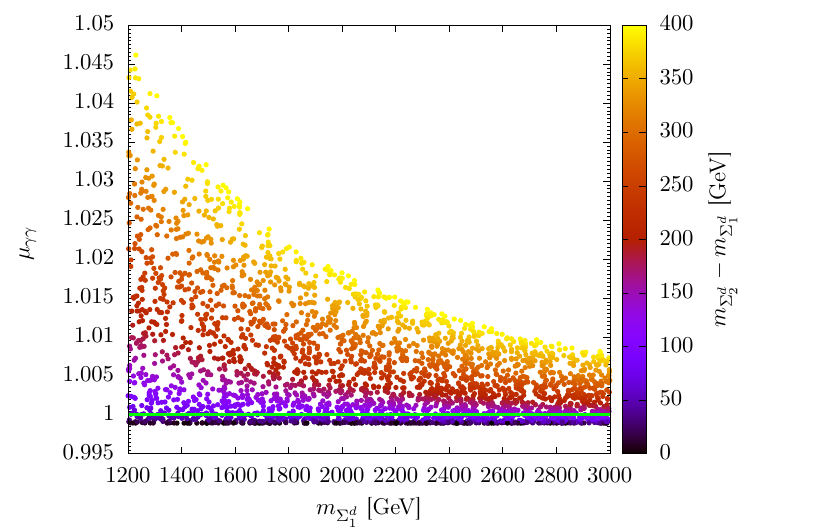} \\
        (A) \hspace{7cm} (B)\\
		\caption{Panel (A): Variation of the loop amplitude for the Higgs diphoton decay width arising solely from LQ contributions, given by $\mathcal{A}^{\Sigma^d}_{\gamma\gamma}\equiv \sum_{\Sigma^d} N^C_{{\Sigma^d}} \,Q_{\Sigma^d}^2\, g_{{\Sigma^d}}\, F_0(\tau_{\Sigma^d})$.
Panel (B): Dependence of the Higgs–to–diphoton signal strength on the LQ mass. The colour bar represents the mass splitting between the two down-type LQ states, while the horizontal green line indicates the SM prediction ($\mu=1$).}
		\label{fig:haa}
	\end{figure}

In the left panel of Fig.~\ref{fig:haa}, we show the contribution of leptoquarks to the Higgs diphoton amplitude. Although this contribution is relatively small, it is positive. For comparison, the amplitude of the $W^\pm$ loop is approximately $8.326$, while the fermionic loop contribution is $(0.048 - 0.053 \,i)$. Consequently, the constructive interference between the LQ and $W^\pm$ loops can enhance the Higgs diphoton decay width in the presence of LQs.  
When the LQ masses are increased, the amplitude $\mathcal{A}^{\Sigma^d}_{\gamma\gamma}$ decreases due to propagator suppression in the loop. Similarly, when the mass gap between LQs is decreased, it also suppresses the LQ contribution, since the cubic coupling connecting the SM Higgs to the LQs is directly proportional to this mass gap.

In the right panel of Fig.~\ref{fig:haa}, we show the Higgs diphoton signal strength in the case of maximal mixing, $\theta_{\text{max}} \simeq \pi/4$. The predicted signal strength $(\mu_{\gamma\gamma})$ exhibits the same qualitative behaviour as the amplitude $\mathcal{A}^{\Sigma^d}_{\gamma\gamma}$. We note that the SM gauge boson and fermion loop contributions will provide an overall constant shift. We find that the Higgs to diphoton signal strength remains well within the experimental limits for our model. In addition, we observe that the constraints extracted from $h \rightarrow \gamma\gamma$ signal strength are much weaker compared to the $\Delta T$ bound discussed in the previous subsection.
Note that as the VLLs do not couple to the SM Higgs in our model, they do not contribute to the $h \rightarrow \gamma\gamma$ amplitude.

\paragraph{Flavour constraints --} Due to their Yukawa couplings with SM quarks, LQs can, in general, induce contributions to neutral meson oscillations such as $K^0\!-\!\bar{K}^0$, $B_d^0\!-\!\bar{B}_d^0$, and $B_s^0\!-\!\bar{B}_s^0$ mixing \cite{Kumar:2016omp}. To avoid such effects, in our model, we restrict the LQs to couple only to third-generation quarks. This choice ensures that no leading-order contribution to meson oscillations arises from LQ interactions. Moreover, since the LQs in our model are odd under the $Z_2$ symmetry, they neither behave as diquarks nor couple to SM leptons. Consequently, flavour-violating leptonic meson decays do not impose constraints on our model, in contrast to flavoured LQ scenarios where such couplings are present \cite{Varzielas:2023qlb}, making our model significantly less constrained by low-energy flavour physics.

\section{Vector like lepton mass} \label{sec:vlmsplit}

\begin{figure}[H]
	\centering
	\includegraphics[angle=0,width=0.4\textwidth]{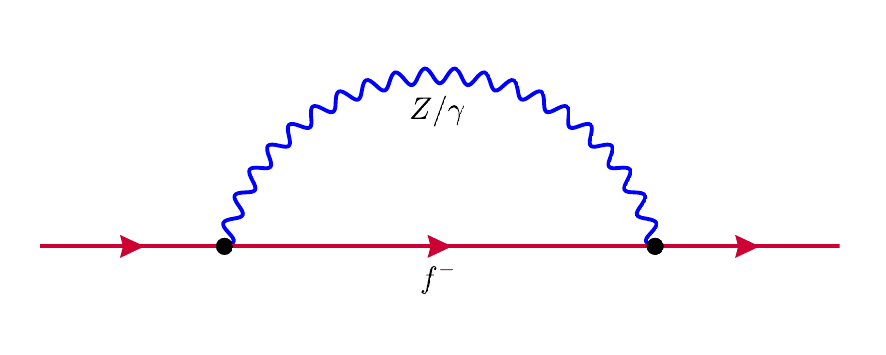} 
	\includegraphics[angle=0,width=0.4\textwidth]{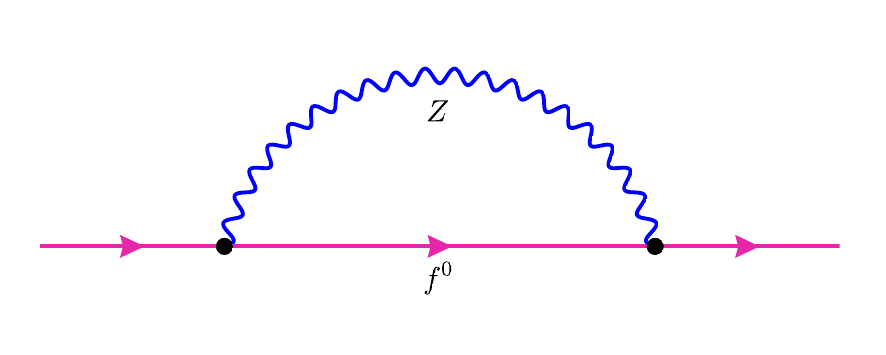}\\
	(A) \hspace{6cm} (B)\\
	\includegraphics[angle=0,width=0.4\textwidth]{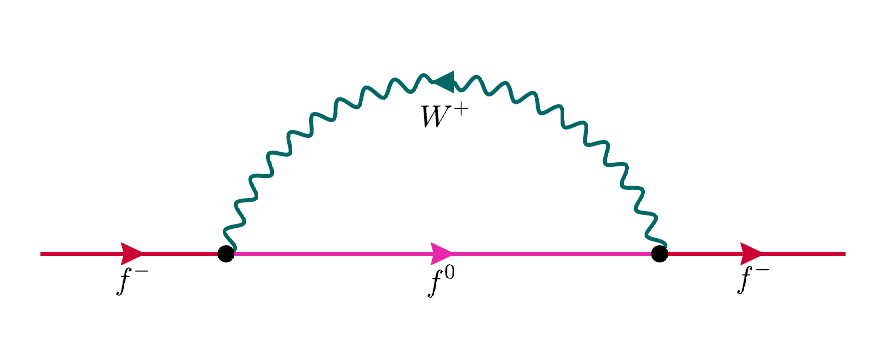} 
	\includegraphics[angle=0,width=0.4\textwidth]{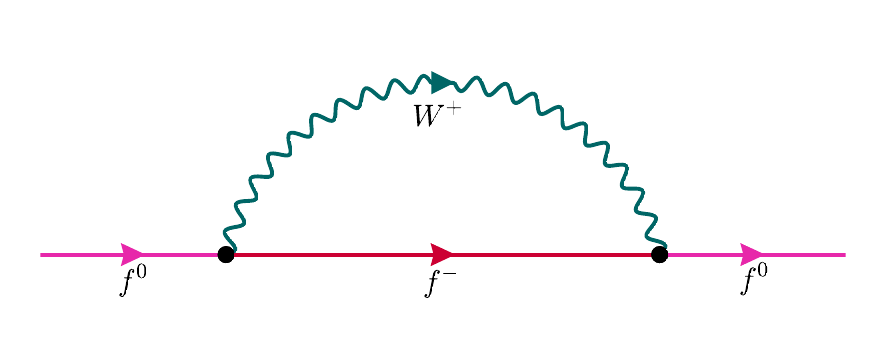}\\
	(C) \hspace{6cm} (D)\\
	\includegraphics[angle=0,width=0.48\textwidth]{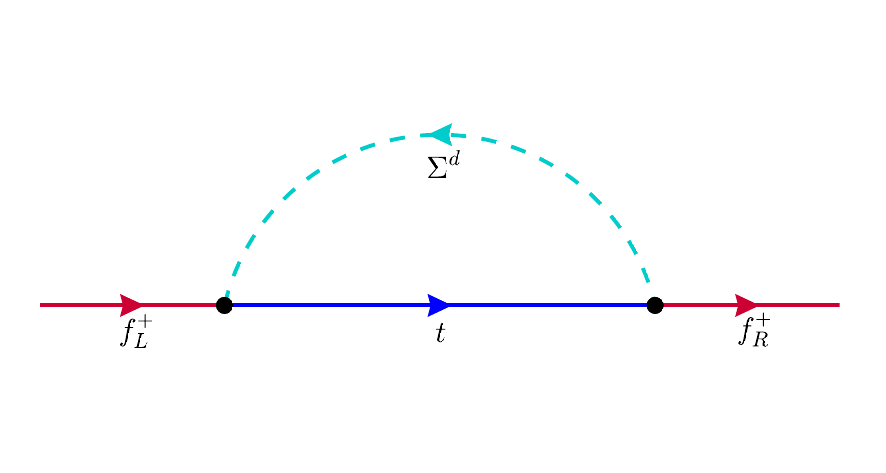}\\
	(E) \\
	\caption{Feynman diagrams contributing to mass gap of charged and neutral component of the vector-like doublet  }
	\label{fig:diracmass}
\end{figure}

At tree level, the neutral and charged components of the VLL doublet are mass-degenerate, with their common mass given by the Dirac mass term in the Lagrangian (see Eq.~\ref{eq:yukawa}). However, as discussed in Sec.~\ref{sec:pureVLL}, this degeneracy is lifted at the quantum level as the electroweak gauge boson loops generate a radiative mass splitting between the charged and neutral VLL states. The relevant Feynman diagrams that lead to the mass splittings are shown in Fig.~\ref{fig:diracmass}.      
In our model, the presence of scalar leptoquarks induces additional one-loop 
contributions to the masses of the charged and neutral fermionic components, $f^{\pm}$ and $f^{0}$, as shown in Fig.~\ref{fig:diracmass}. These additional one-loop contributions enhance the mass splitting between the charged and neutral states from the usual expected contributions of electroweak gauge bosons alone. The final expression for the Dirac mass splitting between charged and neutral fermions is given by 
\begin{eqnarray}
 \label{chargedvllmassgap}
	\delta m_D=\delta m_{EW}+3m_f y_a y^\dagger_c \sum_{i,j=1}^{2} U_{RS}(i,i)U^\dagger_{RS}(i,j)(B_1(m^2_f, m^2_t, m^2_{\Sigma^d_i}) - B_0(m^2_f, m^2_t, m^2_{\Sigma^d_i})) \,\,\, ,
\end{eqnarray} 
where $\delta m_{EW}$ denotes the sole contribution
from the electroweak gauge-boson loops. Its expression coincides with $\delta m_{\pm}$ defined in Eq.~\ref{fig:vllpm}. The loop functions $B_0$ and $B_1$ 
appearing in the expression are the standard 
Passarino--Veltman scalar and tensor integrals~\cite{Passarino:1978jh}, which parameterize the one-loop corrections.

\begin{figure}[H]
	\centering
	\includegraphics[angle=0,width=0.5\textwidth]{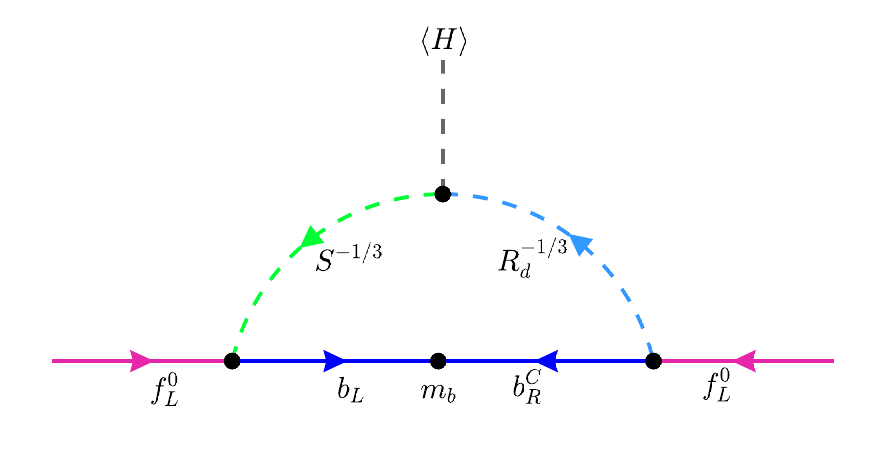} 

	\caption{Feynman diagram contributing to the Majorana mass term of the neutral component  of the vector-like doublet  }
	\label{fig:mass2}
	\end{figure}

In addition to the Dirac mass splitting, the neutral VLF receives a small Majorana mass at one loop due to the presence of both fermion--number--violating and fermion--number--conserving interactions in Eq.~\ref{eq:yukawa}. The relevant Feynman diagram is shown in Fig.~\ref{fig:mass2}. This radiatively generated Majorana mass leads to a further splitting of the Dirac fermion into two nearly-degenerate pseudo-Dirac states.
The induced one-loop Majorana mass is then given by
\begin{equation}
 \label{eqn:majoranamass}
	m_L \sim 6 m_b y_a y^\dagger_b \sum_{i,j=1}^{2} U_{RS}(i,i)U^\dagger_{RS}(i,j) B_0( m^2_b,0, m^2_{\Sigma^d_i}) \,\,\, .
\end{equation} 	
Note that the Majorana mass term is possible only for the left-chiral projection of the neutral VLF, 
since both both fermion--number--violating and fermion--number--conserving vertices (Eq.~\ref{eq:yukawa}) exist only for the left-handed component in the Lagrangian.

Thus, the final mass matrix for the neutral components is given by 
\begin{equation}\label{eqn:neutralmatrix}
	M_{f^0}=\left(\begin{matrix} m_L~~~~m^0_{1~loop}  \\ m^0_{1~loop}~~~~0 \end{matrix}\right)
\end{equation}
where $m^0_{1~loop}$ is the one-loop Dirac mass of the neutral component, which is given by
\begin{equation}
   m^0_{1~loop}=m_f+ \mathrm{Real}\left({\sum}^{f}_{W}+{\sum}^{f}_Z\right)\,\,\,.
\end{equation}
Here ${\sum}^{f}_{W}$ and ${\sum}^{f}_{Z}$ are the VLL self-energy corrections due to the $W$ and $Z$ bosons in the loop, respectively. The generic expression for the self-energy terms is given as
\begin{align}
    {\sum}^f_V=&\dfrac{g^2_V}{16\pi^2}\dfrac{m_f}{m^2_V}\left[A_0(m_V)-A_0(m_f)\right. \nonumber \\+&\left.(5m_V^2-2m_f^2)B_0(m_f,m_V)+(2m_V^2-m_f^2)B_1(m_f,m_V)\right] \,\,\,.
\end{align}
Here $V\in\{W^\pm,Z\}$ and the coupling of VLL with the weak gauge bosons are given by,
\begin{equation}
  g_W=-\dfrac{ig_2}{\sqrt{2}}, \qquad g_W=-\dfrac{i(g_1 \sin{\theta_W+g_2 \cos{\theta_W}})}{2} \,\,\,.  
\end{equation}

The physical mass eigenstates of the two pseudo-Dirac fermions are then given by
\begin{equation}\label{eqn:neutralmass}
    m_{f^0_{1,2}}=\frac{m_L}{2}\pm \sqrt{\frac{m^2_L}{4}+(m^0_{1~loop})^2} \,\,\, .
\end{equation}
We find that the calculations yield negative values for $m_L$, which makes $m_{f^0_1}$ the lighter state. We will identify $f^0_1$ as the DM candidate in our study.

\begin{figure}[H]
		\centering
		\includegraphics[angle=90,width=0.47\textwidth]{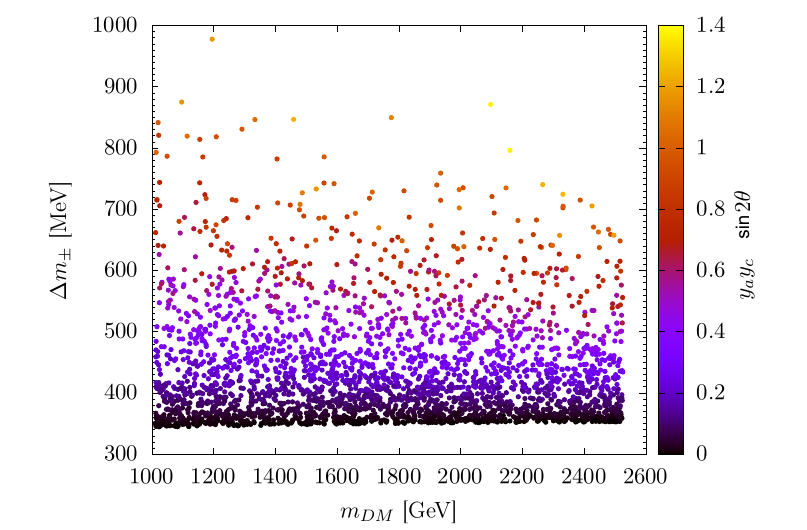} 
		\includegraphics[angle=90,width=0.48\textwidth]{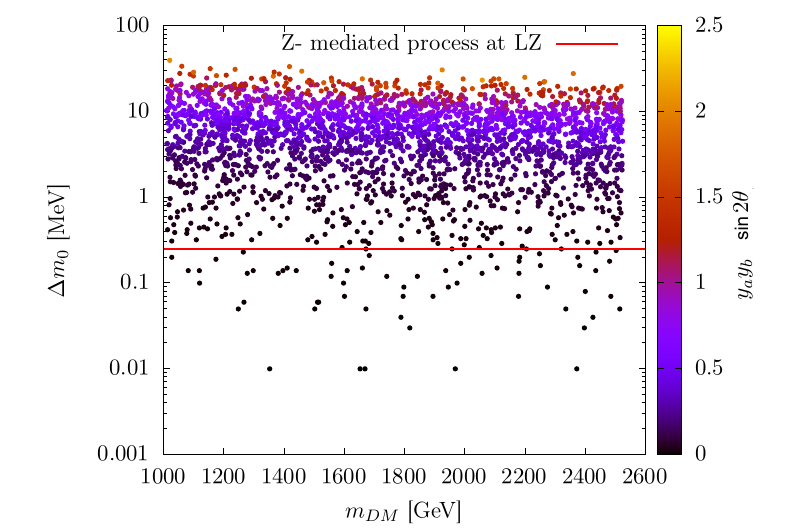}\\
		(A) \hspace{7.2cm} (B)
		\caption{The figure on the left (A) shows the variation in mass splitting between the charged and neutral components of the VLL. Figure (B) on the right shows the variation in mass splitting between the two neutral pseudo-Dirac fermions as a function of the mass of the lightest of the two states.}
		\label{fig:massplots}
	\end{figure}

Next, we carry out a parameter scan over 
$(m_f,\, y_a,\, y_b,\, y_c,\, \sin\theta)$ to study the dependence of the mass splittings between the charged fermion and the neutral pseudo-Dirac 
states. As shown in Fig.~\ref{fig:massplots}(A), the inclusion of the LQs significantly enhances the mass gap. In particular, the splitting between the charged and lighter neutral fermions can
substantially exceed the value of $355~\mathrm{MeV}$, obtained when loop corrections arise solely from electroweak gauge bosons. 

In Fig. \ref{fig:massplots}(B), where we have shown the variation in the mass splitting between the two pseudo-Dirac fermions, we observe that for most of the parameter space the two pseudo-Dirac fermions have a mass splitting above $250$ keV. This mass splitting is crucial in our analysis as it forbids the $Z$-mediated DM-nucleon scattering process at the LUX-ZEPLIN \cite{PRL.131.041002} experiment, which was earlier possible for the degenerate case.  With the coupling of the DM with $Z$ gone, it now becomes easier to evade the stringent DD bounds.

\section{DM Phenomenology in presence of inert LQ}\label{sec:dmpheno}
We have shown in Sec.~\ref{sec:pureVLL} that the pure VLL model encounters significant challenges in satisfying both relic density and direct detection constraints. To illustrate how the inclusion of LQs can improve the DM phenomenology of VLLs, we will now analyze their impact on the relic density, indirect detection, and DD signatures. For this purpose, we perform a parameter space scan within the following ranges. 
    \begin{align}
        1000 ~\text{GeV} \leq  m_{DM}\equiv m_{f^0_1} \leq 2500 ~\text{GeV} \nonumber\\
        350 ~\text{GeV}\leq  m_{\Sigma^d_1}-m_{DM} \leq 1000 ~\text{GeV}\nonumber \\
        0.5 \leq  y_a \leq 1.5, \quad
        0.5 \leq  y_b \leq 1.5,\quad
        0.5 \leq  y_c \leq 1 \nonumber \\
        0^\circ \leq  \theta \leq 45^\circ \nonumber \\
        m_{\Sigma^d_2}-m_{\Sigma^d_1}=128~\text{GeV} \nonumber
    \end{align}
From the study of the pure VLL model, we find that the dark matter (DM) becomes under-abundant for masses below $\sim 1150$ GeV. The inclusion of scalar leptoquarks (LQs) further reduces the relic density due to additional (co)annihilation channels. For this reason, we restrict our analysis to DM masses above $1000$ GeV, where the presence of LQs can positively impact the phenomenology. In addition, the Yukawa couplings $y_{a,b,c}$ are constrained to be above $0.25$ from DD observations. This is because the DD constraint is dependent on the mass splitting $\Delta m_0$ between the two light pseudo-Dirac states, which in turn is proportional to the Yukawa couplings ($\Delta m_0 \propto y_a y_b\sin{2\theta}$) mentioned above. We therefore restrict ourselves to Yukawa couplings greater than $0.5$ in our analysis. Furthermore, we fix the mass gap between the two down-type LQs at $128$ GeV, which lies close to the maximum value allowed by the $\Delta T$ parameter constraint in our model. While in principle this mass splitting can be chosen smaller, doing so enhances the negative one-loop corrections to the LQ quartic couplings \cite{Babu:2010vp}. To disentangle scalar-sector constraints from the DM phenomenology, we therefore fix $m_{\Sigma^d_2}-m_{\Sigma^d_1}$ at a value that is consistent with all current experimental bounds.
For the numerical study, we first implement the model in the publicly available code \texttt{SARAH} \cite{Staub:2013tta} to generate the code for \texttt{SPheno} \cite{POROD20122458} and CalcHEP \cite{Belyaev:2012qa} model files, which will be used by \texttt{MicrOmegas} \cite{Alguero:2023zol}. We have used \texttt{SPheno} to generate the particle mass spectrum. And then finally, with the help of \texttt{MicrOmegas}, we evaluate the DM relic density, DD, and ID cross-sections.
\subsection{Relic Density}In this section, we will discuss the relic density of $f^0_1$, which is the lightest neutral state in the dark sector (DS) and is the DM candidate. To calculate the DM relic density, one needs to solve the Boltzmann equation 
\begin{equation}\label{eqn:boltzmann}
    \frac{dn}{dt} + 3Hn=-\langle\sigma v\rangle_{\rm eff}(n^2-n^2_{eq}) \,\,\, .
\end{equation}
Here, $n$ denotes the total number density of all DS particles, i.e., those that are odd under the discrete $Z_{2}$ symmetry, while $H$ represents the Hubble expansion rate. The quantity $\langle\sigma v\rangle_{\rm eff}$ is the effective thermally averaged annihilation cross section is given by
\begin{eqnarray} 
	\label{eqn:sigveff}
	\langle\sigma v\rangle_{\rm eff}&=& \frac{g^2_0}{g^2_{\rm eff}}\langle\sigma v\rangle_{f_{DM} f_{DM}} + \frac{2 g_0 g_i }{g^2_{\rm eff}}\langle\sigma v\rangle_{f_{DM} \chi_i} \Big(1+\Delta_{\chi_i}\Big)^{\frac{3}{2}}  e^{-x \Delta_{ \chi_i}} \nonumber \\
	&& + \frac{2 g_i g_j}{g^2_{\rm eff}}\langle\sigma v\rangle_{\chi_i \chi_j} \Big(1+\chi_i\Big)^{\frac{3}{2}}\Big(1+\chi_j)^{\frac{3}{2}} e^{-x (\Delta_{\chi_i}+\Delta_{\chi_j})} \, \,\,\, .\hspace{0.5cm}
\end{eqnarray}
Here $\chi \in \{f^0_2,f^{\pm},H^d_{1,2},H^u\} $, $\Delta_{\chi_i} = \frac{(M_{\chi_i}-M_{DM})}{M_{DM}}$ and $g_i$ is the multiplicity of the $i^{th}$ particle.
This modification relative to the usual $\langle\sigma v\rangle$ is required because we are evolving the number density of the entire DS sector, rather than just that of the DM candidate. Since all heavier DS states are expected to have decayed into the stable DM particle well before the present epoch, the final DM relic abundance is simply determined by the total DS number density. A detailed account of this method can be found in Ref.~\cite{PRD.43.3191}.

\begin{figure}[H]
	\centering
	\includegraphics[angle=0,width=0.3\textwidth]{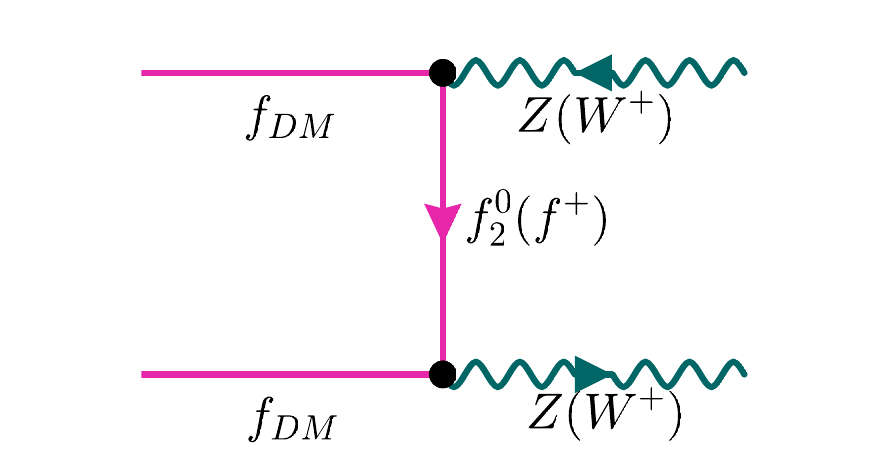} 
	\includegraphics[angle=0,width=0.3\textwidth]{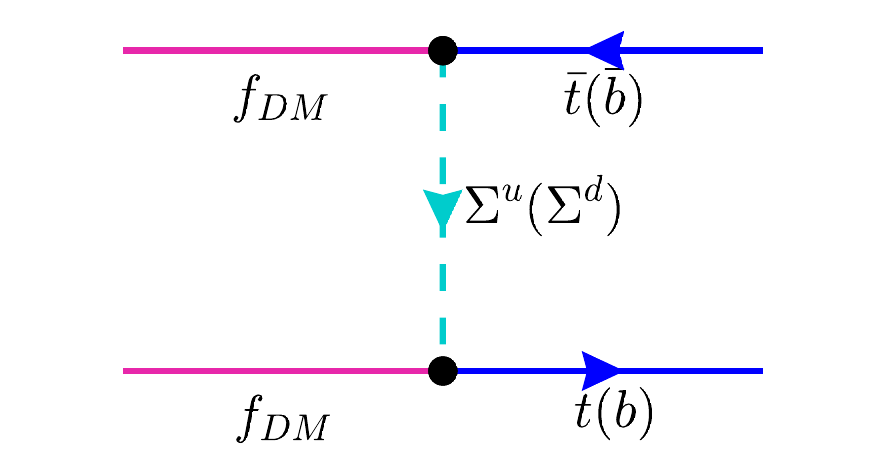}\\
	(A) \hspace{4.3cm} (B)
	\caption{Illustrating the Feynman diagrams for DM self-annihilation processes. These will be referred to as \textbf{Type-I} channels.}
	\label{fig:typeA}
\end{figure}

Next, we consider the processes that contribute to 
$\langle\sigma v\rangle_{\rm eff}$ and indicate their corresponding Feynman diagrams. The first term in Eq.~\ref{eqn:sigveff} originates from DM self-annihilation, shown in Fig.~\ref{fig:typeA}. For simplicity, we denote these as \textbf{Type-I} processes. In these channels, the final states consist of SM gauge bosons or top quarks, mediated through $t$-channel exchange of LQs and VLLs. Since such $t$-channel contributions are strongly suppressed when both the DM and the LQs are heavy, these processes constitute only a subdominant contribution to $\langle\sigma v\rangle_{\rm eff}$.

\begin{figure}[H]
	\centering
	\includegraphics[angle=0,width=0.3\textwidth]{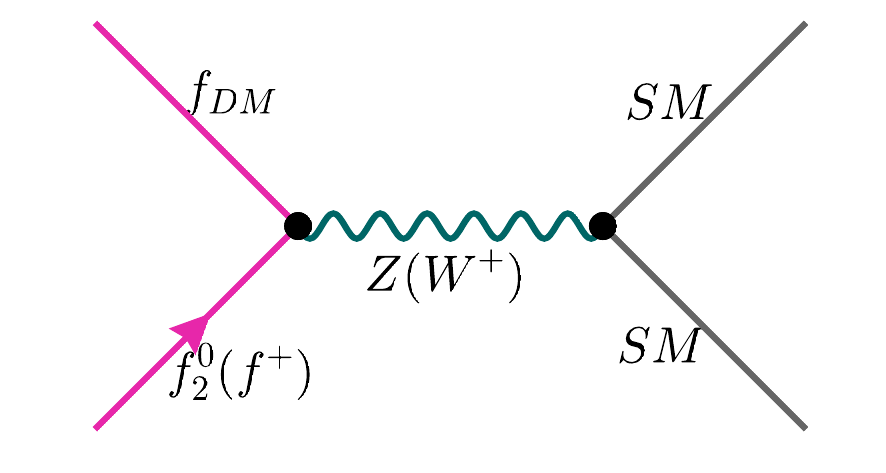} 
	\includegraphics[angle=0,width=0.3\textwidth]{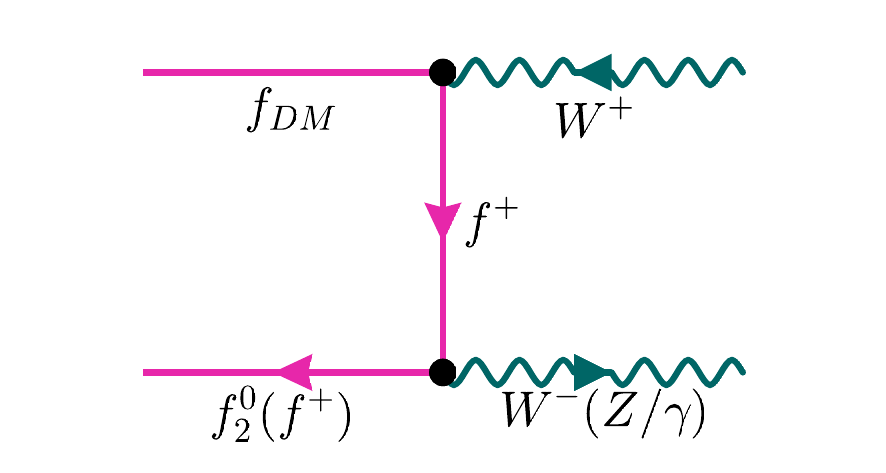}\\
	(A) \hspace{4.3cm} (B)\\
	\includegraphics[angle=0,width=0.3\textwidth]{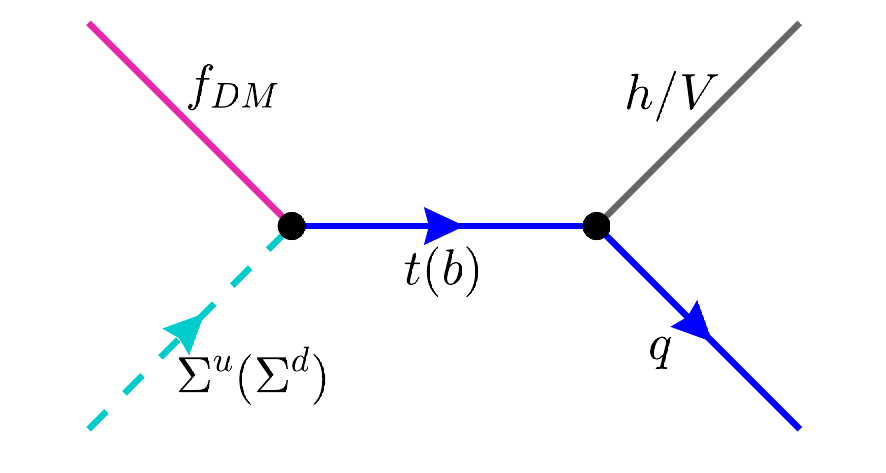} 
	\includegraphics[angle=0,width=0.3\textwidth]{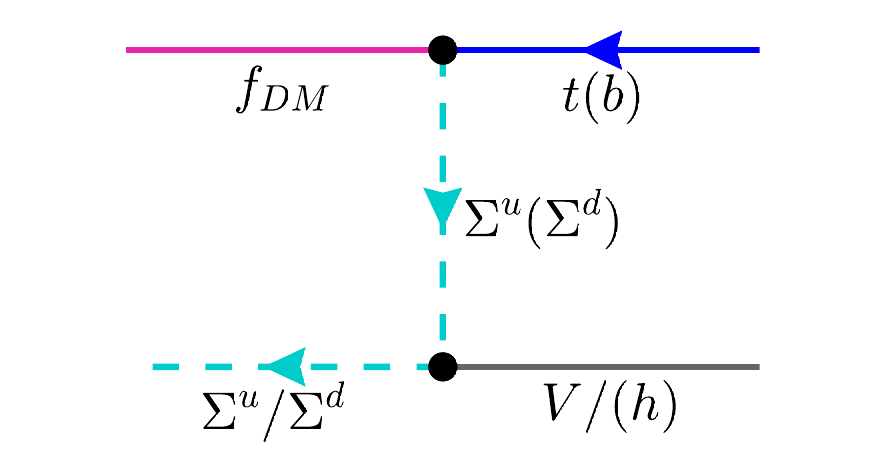}\\
    (C) \hspace{4.3cm} (D)\\
	\caption{Illustrating the Feynman diagrams for DM co-annihilation processes where only one of the incoming particles is DM. These will be referred to as \textbf{Type-II} channels.}
	\label{fig:typeB}
\end{figure}	

The second term in the expression for $\langle\sigma v\rangle_{\rm eff}$ arises from co-annihilation processes between DM and other DS particles. The corresponding Feynman diagrams, shown in Fig.~\ref{fig:typeB}, are denoted as \textbf{Type-II} processes. These channels include both $s$-channel (mediators are SM particles) and $t$-channel (mediators are heavy DS states) contributions. The overall contribution of these diagrams is subject to the Boltzmann suppression factor appearing in Eq.~\ref{eqn:sigveff}, which implies that co-annihilation channels become less relevant when the mass splitting between the two incoming particles is large.  

In the present scenario, the vector-like leptons are nearly mass-degenerate with the DM candidate, with mass splittings of only a few MeV. Consequently, processes (A) and (B) dominate among the Type-II contributions. Although diagram (C) is more strongly Boltzmann-suppressed relative to (A) and (B), its mediator is an SM top or bottom quark, which results in moderate kinematic suppression. Furthermore, the presence of three coloured states for the $\Sigma^{u/d}$ particles, together with the choice of a relatively large coupling at the incoming vertex, enhances the cross section, making diagram (C) a significant contributor to $\langle\sigma v\rangle_{\rm eff}$ despite the suppression.
\begin{figure}[h!]
	\centering
	\includegraphics[angle=0,width=0.3\textwidth]{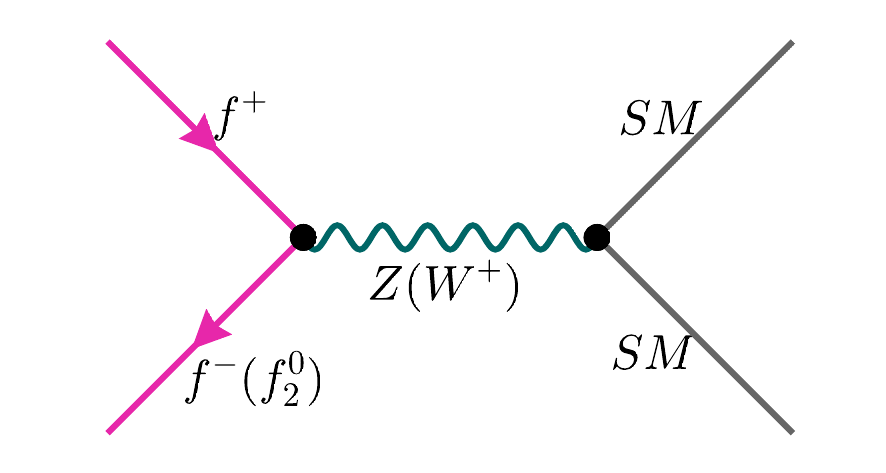} 
	\includegraphics[angle=0,width=0.3\textwidth]{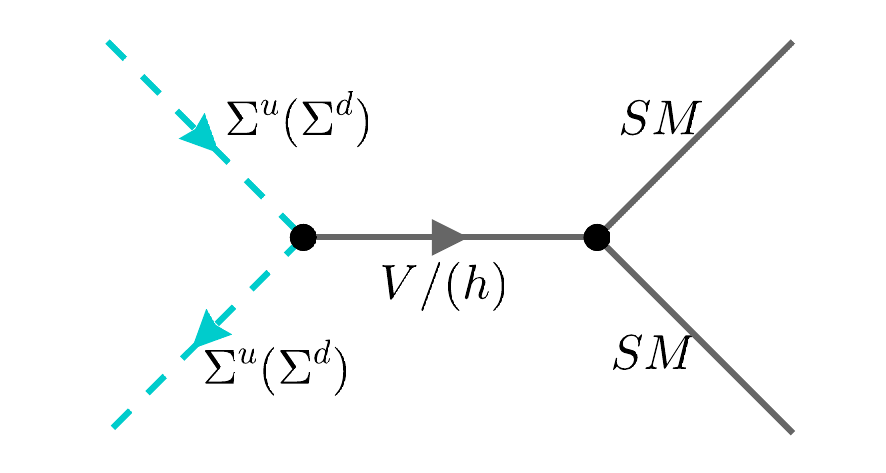}\\
	(A) \hspace{4.3cm} (B)\\
	\includegraphics[angle=0,width=0.3\textwidth]{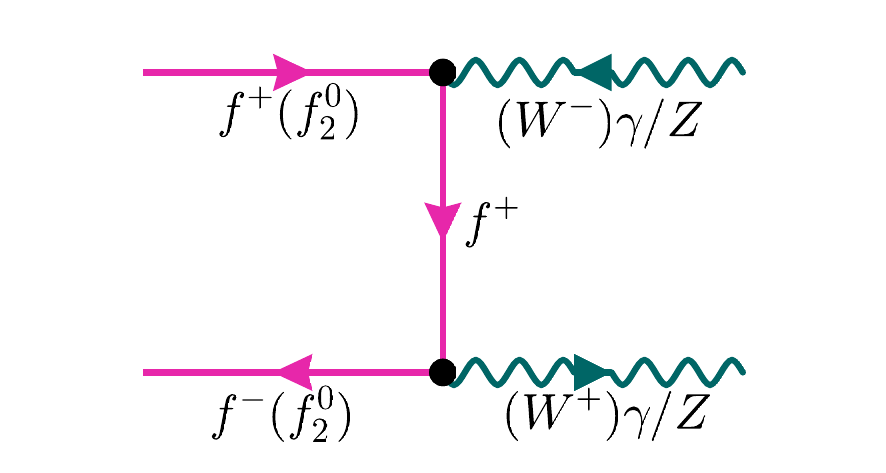} 
	\includegraphics[angle=0,width=0.3\textwidth]{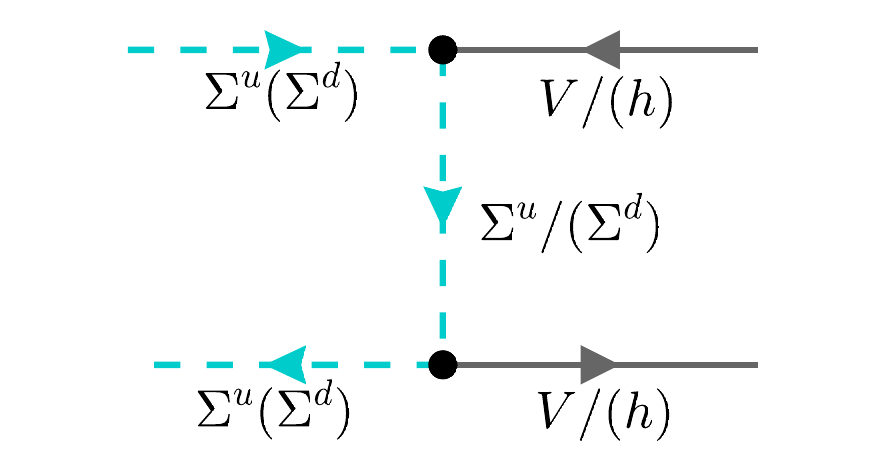}
	\includegraphics[angle=0,width=0.3\textwidth]{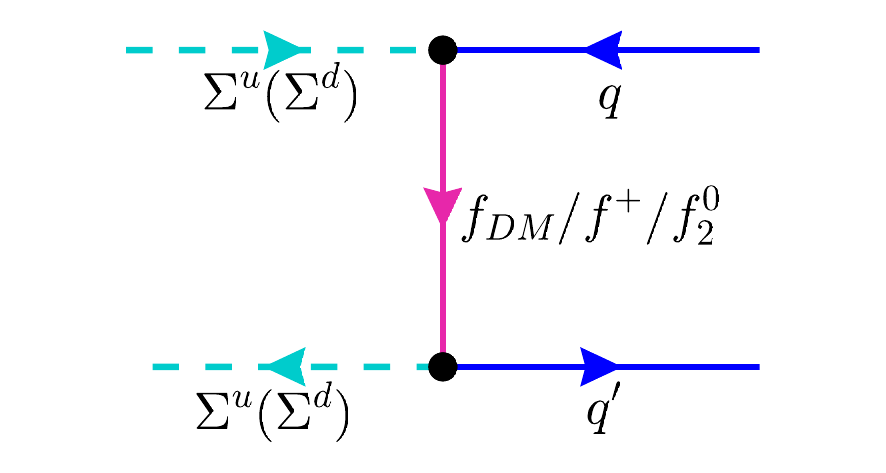}\\
	(C) \hspace{4.3cm} (D)\hspace{4.3cm} (E) 
	\caption{Illustrating the Feynman diagrams for annihilation processes between DS particles where none of the incoming particles is DM. These will be referred to as \textbf{Type-III} channels. }
	\label{fig:typeC}
\end{figure}
Finally, the last term in the expression for $\langle\sigma v\rangle_{\rm eff}$ originates from annihilation processes involving only the heavier dark sector particles, without the DM in the initial state. Since the DM and other DS states remain in chemical equilibrium during freeze-out, any change in the number density of these heavier particles feeds back into the DM abundance through the coupled Boltzmann equations. The corresponding Feynman diagrams for these processes are shown in Fig.~\ref{fig:typeC}, and will be denoted as \textbf{Type-III} processes. We observe that the Type-III processes lead to cross-sections that are greater than that of Type-I processes. This is because the Type-I process is mediated by the heavy BSM propagators in the $t$ channel, leading to a stronger suppression compared to the Type-III processes with SM gauge boson mediators. In addition, there is further interplay of the interactions via gauge coupling and BSM Yukawa couplings that favour the Type-III processes over Type-I. Thus the contribution to $\langle\sigma v\rangle_{\rm eff}$ from (co)annihilation of dark sector particles other than DM can be significantly high even with the extra Boltzmann suppression, this makes Type-III processes the second leading contributor to $\langle\sigma v\rangle_{\rm eff}$ next to Type-II processes which contributes the most.
 
\begin{figure}[H]
\centering
\includegraphics[width=0.47\textwidth]{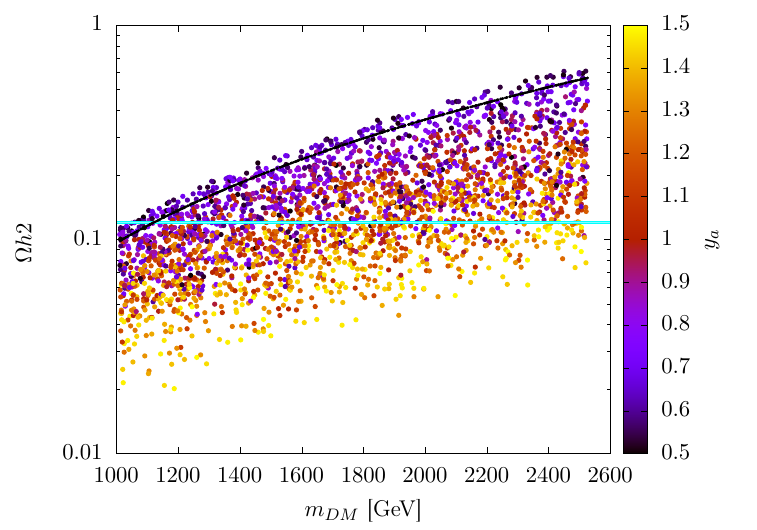}
\includegraphics[width=0.50\textwidth]{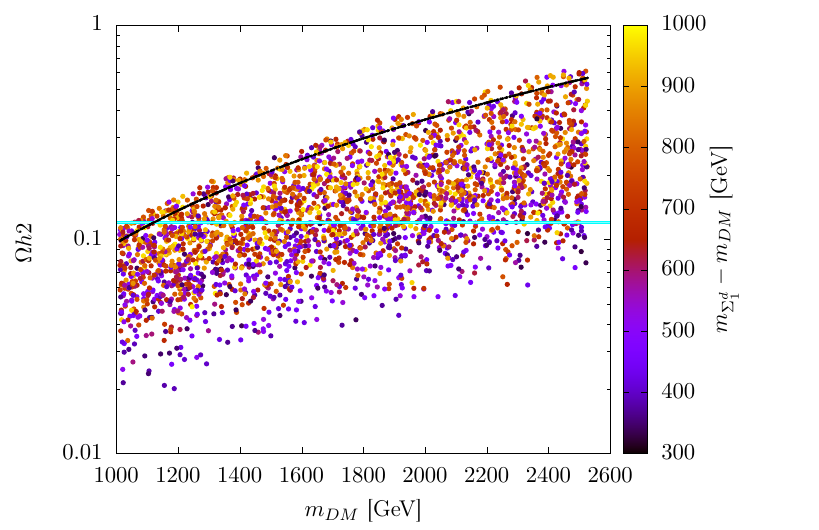}\\
(A) \hspace{7.2cm} (B)\\
\caption{Scatter plots of the DM relic density as a function of its mass. In panel (A), the color bar denotes the coupling $y_a$; in panel (B), it shows the mass gap between the DM and the lightest down-type LQ. The solid black curve corresponds to the relic density in the pure VLL model, while the cyan line indicates the observed DM relic density, $0.12 \pm 0.001$.}
\label{fig:relic1}
\end{figure}
We now turn to the phenomenological implications of the model parameters on the DM relic density, highlighting how they shape the freeze-out dynamics. The key features are shown in Fig.~\ref{fig:relic1}.  In Fig.~\ref{fig:relic1}(A), we observe that the relic density decreases as the coupling $y_a$ is increased (a similar behaviour is observed for $y_b$ and $y_c$, not shown here). This behavior is expected, since the strength of the LQ--VLL interactions is controlled by these couplings. For small couplings, the relic density approaches the value obtained in the pure VLL scenario, where the LQ interactions are effectively absent.  Panel (B) of Fig.~\ref{fig:relic1} shows that the relic density is also sensitive to the mass gap between the lightest leptoquark, $\Sigma^d_1$, and the DM candidate. This dependence arises because the mass gap governs the strength of the Boltzmann suppression for LQ-(co)annihilation processes, where a larger mass splitting leads to stronger suppression, and hence to a smaller impact on the relic density. By varying both the LQ--VLL couplings and the LQ--DM mass gap, the observed relic abundance can be reproduced in our model.
\begin{figure}[H]
\centering
\includegraphics[width=0.49\textwidth]{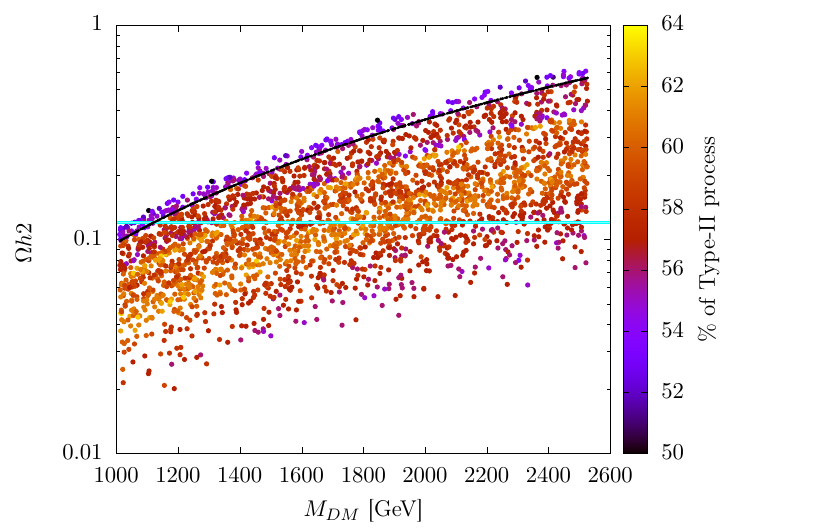}
\includegraphics[width=0.49\textwidth]{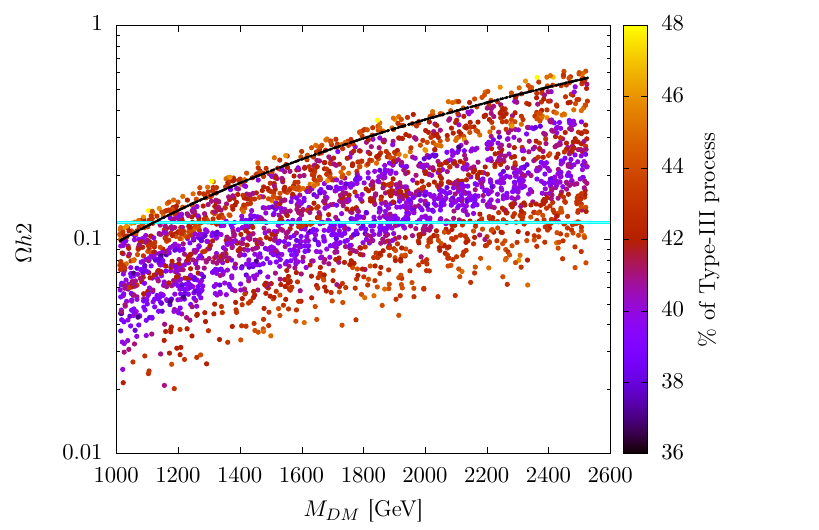}\\
(A) \hspace{7.2cm} (B)\\
\includegraphics[width=0.49\textwidth]{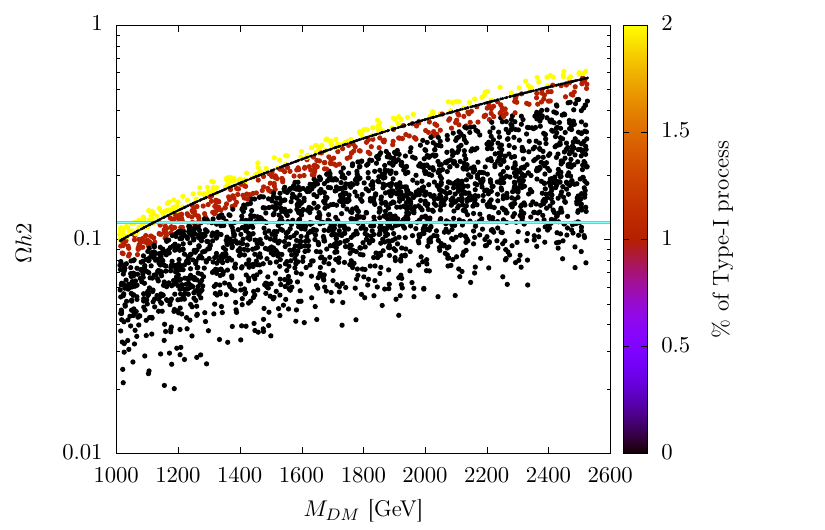}\\
(C)
\caption{The panel (A), (B) and (C) shows the fractional contribution of the Type-II, III and I processes to the total $\langle\sigma v\rangle_{\rm eff}$ in the Boltzmann equation. The solid black curve corresponds to the relic density in the pure VLL model, while the cyan line indicates the observed DM relic density, $0.12 \pm 0.001$.}
\label{fig:relic2}
\end{figure}
 Fig.~\ref{fig:relic2}(A) shows how the presence of LQs modifies the overall effective cross section $\langle\sigma v\rangle_{\rm eff}$ compared to the pure VLL case. In the pure VLL scenario, the DS mass spectrum is highly compressed, so annihilation and co-annihilation processes contribute comparably, with the Boltzmann factor $\sim 1$ implying negligible suppression. Consequently, in Fig.~\ref{fig:relic2}(A) Type-II processes account for roughly $50-64\%$ of the total cross section while from Fig.~\ref{fig:relic2}(B) we can see the Type-III processes contributes between $36-48\%$, this indicates a significant departure from the pure VLL case as with the inclusion of LQs the contribution from annihilation of pseudo Dirac DM in negligible ($\leq2\%$). The main reason behind this is pseudo Dirac DM can only annihilate via heavy BSM particle as mediator, thus it is highly suppressed by their propagators.    

\subsection{Indirect Detection}
Dark matter can also be probed through indirect detection experiments such as H.E.S.S. \cite{PRL.129.111101}, FermiLAT and MAGIC \cite{MAGIC_2016}, which search for excess cosmic-ray or gamma-ray signals arising from DM annihilation into Standard Model final states, e.g., $W^+W^-$, $ZZ$, $b\overline{b}$, and $\tau^+\tau^-$.
\begin{figure}[htb]
	\centering
	\includegraphics[angle=0,width=0.6\textwidth]{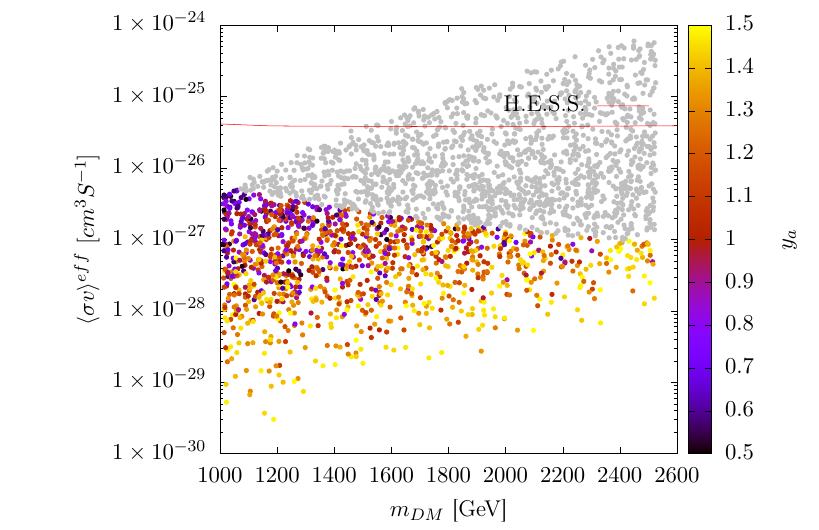}
	\caption{Scatter plot of the velocity-weighted annihilation cross-section $\langle\sigma v\rangle$ as a function of the DM mass for annihilation into $W^+W^-$. The color bar denotes the coupling $y_a$. The cross-section is scaled by the square of the fractional relic abundance $f^0_1$ that the DM can account for. Gray points indicate parameter regions excluded by relic density constraints. The red curve represents the indirect detection bound given by H.E.S.S. \cite{PRL.129.111101}.  }
	\label{fig:id}
	\end{figure}
 In our model, the $W^+W^-$ final state provides the dominant ID signature, and H.E.S.S. \cite{PRL.129.111101} observations particularly constrain this channel. In Fig.~\ref{fig:id}, we show the effective velocity-weighted annihilation cross-section, defined as ${\langle \sigma v\rangle}_{eff}=\left(\frac{\Omega h^2}{0.12}\right)^2\times\langle \sigma v\rangle$ plotted as a function of the DM mass, with the color bar representing the DM–LQ coupling $y_a$. The gray points denote regions of parameter space where DM is overabundant, and excluded by the Planck measurement of the relic density \cite{Planck:2018vyg}. The colored points lying below the gray points correspond to regions consistent with relic density bounds, and we find that this entire viable parameter space also satisfies the indirect detection limits set by H.E.S.S. Thus, the model has a large region of parameter space that remains unconstrained by current indirect searches.
\subsection{Direct Detection}
\begin{figure}[H]
	\centering
	\includegraphics[angle=0,width=0.3\textwidth]{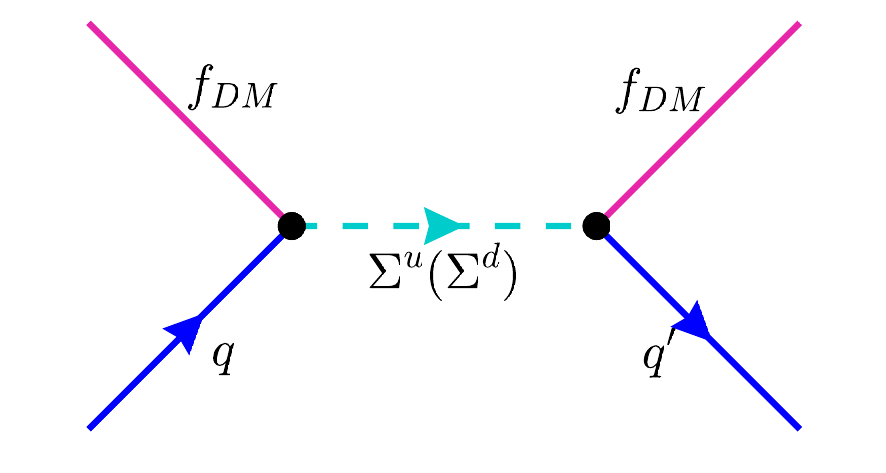}
	\includegraphics[angle=0,width=0.3\textwidth]{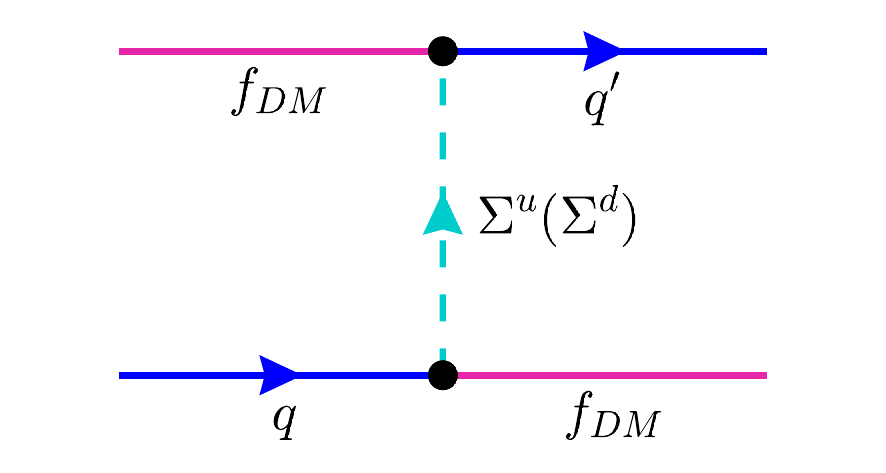}\\
	(A) \hspace{4.3cm} (B)\\
	\caption{Tree level Feynman diagrams contributing to direct detection cross-section. }
	\label{fig:dd}
\end{figure}

In this section, we discuss the direct detection prospects of the DM candidate in our model. Since the DM is a pseudo-Dirac fermion, the tree-level $Z$-mediated DM–nucleon scattering process is absent, helping us evade the stringent $Z$-exchange constraints. However, the presence of LQs introduces new channels for DD. In particular, DM–nucleon scattering can now proceed via LQ-mediated processes, as illustrated in Fig.~\ref{fig:dd}.

The effective spin-independent direct detection cross-section is defined as  
\begin{equation}
\sigma^{\rm eff}_{\rm SI} \;=\; \left(\frac{\Omega h^2}{0.12}\right)\, \sigma_{\rm SI} \,\,\, .
\end{equation}  
\begin{figure}[h]
	\centering
	\includegraphics[angle=0,width=0.46\textwidth]{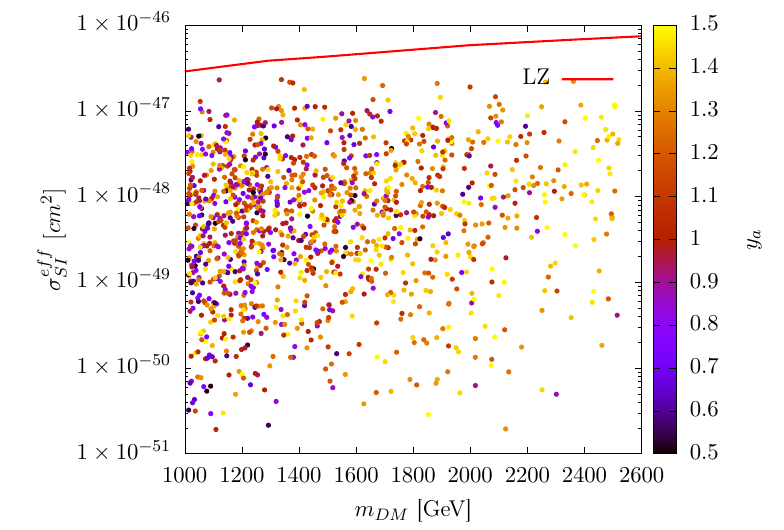}
	\includegraphics[angle=0,width=0.485\textwidth]{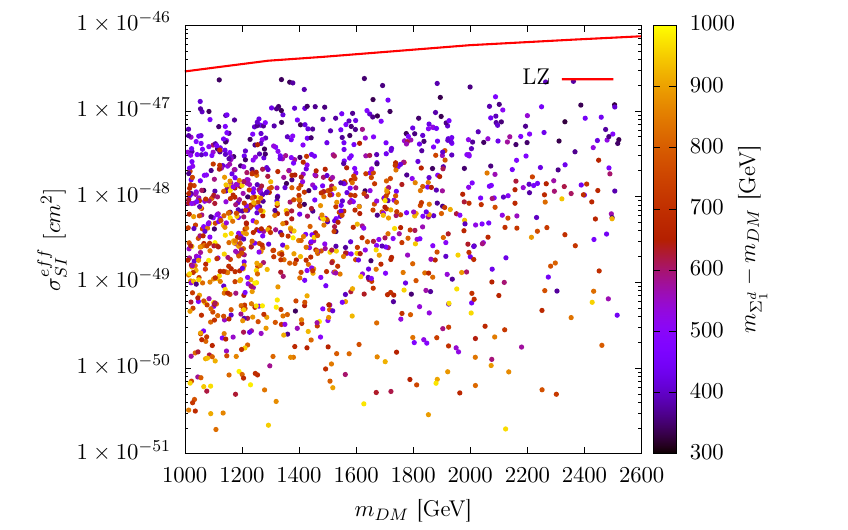}\\
	(A) \hspace{7.2cm} (B)\\
	\caption{Scatter plot of left of spin-independent direct detection cross-section of DM with its mass. The color bar on left plot (A) represents the coupling $y_a$ and on right plot (B) the color bar represents the mass gap between the lightest LQ and DM. The direct detection has been scaled with the fraction of total observed DM relic which  $f^0_1$ can satisfy. The red curve represents the bound on spin-independent direct detection cross-section given by LUX-ZEPLIN \cite{LZ:2024zvo}.  }
	\label{fig:LZ}
\end{figure}
 In Fig.~\ref{fig:LZ}, we present the variation of 
 $\sigma^{\rm eff}_{\rm SI}$ with relevant parameters, where all the displayed points are consistent with both the relic density bounds from \textit{Planck} and the indirect detection limits discussed in the previous section. Panel (A) of Fig.~\ref{fig:LZ} shows the correlation between the DM mass and the required coupling strength. For heavier DM masses, we find that larger values of the DM--LQ coupling are necessary. We observe that as the DM mass increases, the annihilation cross-section decreases, leading to an overabundant relic density. To counter this effect and obtain a consistent observed relic abundance, stronger couplings are required to enhance the annihilation rate. This leads to an allowed parameter space where, for heavier DM, we find regions of larger coupling values.  Panel (B) of Fig.~\ref{fig:LZ} shows the dependence of the effective DD cross-section on the mass gap between the DM and the lightest leptoquark. We observe a clear suppression of $\sigma^{\rm eff}_{\rm SI}$ as the mass gap increases. This is a direct consequence of the exchange of a heavy mediator. As the LQ mediates DM--nucleon scattering, an increase in its mass suppresses the scattering amplitude, thereby lowering the effective cross-section.

To place our predictions in context, we consider results of 
 LZ (4.2t $\times$ y)~\cite{LZ:2024zvo} experiment which puts the stringent bound on DD cross-section of WIMP DM. We find that the entire parameter space consistent with relic density and ID constraints remains below the present exclusion limits. In particular, the predicted cross-sections are well below the current sensitivity. However, a substantial part of this parameter space lies within reach of next-generation detectors such DARWIN (200t $\times$ y)~\cite{DARWIN:2016hyl}, LZ (15.3t $\times$ y)~\cite{LZ:2019sgr}, XENONnT (20t $\times$ y)~\cite{XENON:2020kmp} as they can probe DD cross-section of\, $\sim10^{-47}$ $\mathrm{cm^2}$ suggesting that upcoming experiments will provide a decisive probe of the model’s DD predictions.  

\section{Collider Signatures}

In this section, we briefly discuss the collider signals for our model. The collider phenomenology of our model in the presence of vector-like leptons and scalar leptoquarks, all stabilised by a discrete $Z_2$ symmetry can lead to distinct and interesting signals that are already under inspection by the CMS and ATLAS Collaborations. As both VLLs and LQs are $Z_2$-odd, they must be produced in pairs at the LHC, and their decays necessarily involve the stable neutral VLL component that serves as the DM. This leads to a generic prediction of missing transverse energy ($E_T^{\text{miss}}$) in collider events. Note that we have LQs carrying both $Q_{em} =2/3$ and $Q_{em} =1/3$ charge, and we have mainly discussed the role of the $Q_{em} =1/3$ charged state in the DM analysis. However, the collider signature will comprise of both the charged states as they can be produced independently or associatively in pairs.
\paragraph{Scalar Leptoquark Production --}
The scalar LQs will be produced at the LHC in pairs through both QCD and electroweak subprocesses, with the QCD processes dominating wherever they contribute. The following production modes will be relevant:
\begin{equation}
pp \to \Sigma^d_1 \Sigma^{d*}_1 \, , \, \Sigma^d_2 \Sigma^{d*}_2 \, , \, \Sigma^u \Sigma^{u*} \, , \, \Sigma^u \Sigma^{d*}_{1/2} \,\,\,.
\end{equation}
where the QCD subprocesses will dominate the first three production processes, while the last associated production process occurs through the exchange of the $W$ gauge boson.

Being $Z_2$-odd, their decays must involve a DS particle:
\begin{equation}
\Sigma^i \to q + f^0_{1,2} \,\, , \qquad \Sigma^i \to q' + f^\pm, \qquad \Sigma^u \to \Sigma^d + W 
\end{equation}
where $i\equiv u,d$. The charged VLL will further decay into the DM and leptonic or jet final states (the hardness of the leptonic and jet final states will depend on the mass gap between $f^\pm$ and $f^0_{1,2}$). As the mass splittings are at the level of few hundred MeVs, We expect these to be very soft, and can be easily missed in the detector. 
This gives rise to final states such as multijets + $E_T^{\text{miss}}$, $n\, \ell + m \, j + E_T^{\text{miss}}$ in association with soft leptons or jets. Note that we have only considered the LQ couplings to third-generation quarks and therefore the quark appearing in the decay of the LQs is either the top quark or the bottom quark.

Standard ATLAS and CMS searches for first generation LQs in the $\ell \ell jj$ final states exclude scalar LQs up to $\sim 1.1$--$1.15$ TeV under the assumption of $100\%$ branching to $\ell q$~\cite{ATLAS:2019qpq,CMS:2019pcx}. More recently, non-resonant effects in high-mass dilepton tails have extended sensitivity to $\sim 4$--$5$ TeV for sizable couplings~\cite{CMS:2025nonresLQ,ATLAS:2025LQtail}. In our model, the $Z_2$ odd nature and its invisible decays reduce the branching fractions to the $\ell q$ mode, weakening these bounds significantly. In addition, we work in the scenario where the LQs couple only to the third-generation quarks. The actual constraints are likely to mimic the bounds of those of the scalar quarks (stop and sbottom squarks) in supersymmetry, as pointed out earlier in Sec.~\ref{sec:const}. 

\paragraph{Vector-Like Lepton Production--}

Similarly, as in the case of the LQ production, the VLLs will also be produced in pairs, albeit through the exchange of electroweak gauge bosons only. The primary production channels are:
\begin{equation}
pp \to f^+ f^- , \qquad f^\pm f_2^0 , \qquad f^\pm f^0_{1} , \qquad f^0_2 f^0_{1} \,\,\,.
\end{equation}
Their $Z_2$-odd nature and corresponding mass splitting between the different VLL states ensures decays involving DM:
\begin{equation}
f^\pm \to W^{*} + f^0_{\text{DM}} \,\,\,.
\end{equation}
The resulting signals include soft leptons and jets + $E_T^{\text{miss}}$. These signals will be very similar to supersymmetry searches for sneutrinos and sleptons, with compressed spectra and very small mass splitting with the neutralino DM~\cite{ATLAS:2017oal,ATLAS:2019lng}.
\paragraph{Mixed LQ--VLL Production--}
Some very interesting possibilities in our framework are the associated production processes:
\begin{equation}
pp \to \Sigma^d_{1,2} + f^0_{1,2}, \qquad \Sigma^{u(*)} + f^{-(+)} , \qquad  
\end{equation}
and following the decay channels mentioned above, lead to multilepton + jets + $E_T^{\text{miss}}$ final states. In addition, for very small Yukawa couplings, the charged VLLs may become long-lived, yielding heavy stable charged particles (HSCPs), displaced vertices, or kinked tracks. Current LHC analyses already exclude such states up to $\sim 1.1$--$1.2$ TeV, with HL-LHC expected to extend this reach~\cite{CMS:2023LL,ATLAS:2024LL}.

In summary, the $Z_2$ symmetry leads to pair production and associated production of the DS particles and guarantees $E_T^{\text{miss}}$ in all collider signatures. While standard ATLAS and CMS searches for these types of exotics constrain the mass spectrum using visible decay channels, substantial parameter space becomes viable for their search, particularly in mixed LQ--VLL cascades and suppressed branching scenarios. A dedicated search at the LHC will yield a much-needed bound for such states. We leave a dedicated collider analysis of our model for future work, where we plan to investigate the full available parameter space, consistent with DM observations and existing experimental searches.

\section{Conclusions}\label{sec:concl} In this work, we started by analysing the limitations of the pure $SU(2)$ doublet VLL DM model. We showed that the pure VLL framework proves to be highly restrictive as a dark matter candidate. In the pure VLL scenario, the relic density requirement for the neutral stable fermion can only be met in a finely tuned mass region near $\sim 1200$ GeV, and even then, the model is effectively excluded by stringent direct detection limits due to unsuppressed $Z$-mediated DM–nucleon scattering. This highlights the necessity of extending the minimal VLL setup to achieve a phenomenologically viable dark matter scenario.
To evade these issues, we introduced two scalar LQs: an $SU(2)_L$ doublet and a singlet. The addition of LQs provides new (co)annihilation channels to the DS particles via their Yukawa interaction with the VLL doublet and helps satisfy the DM relic density bound for DM mass above $1200$ GeV. The inclusion of LQ also splits the neutral VLL component into two pseudo-Dirac states, prohibiting the DM--nucleon $Z$ mediated tree-level process at DD experiments, making it possible to satisfy the DD bounds. We observe that this alternate mechanism of enabling a viable VLL DM helps open up a large parameter space of the VLL sector for satisfying the correct relic abundance. We also provide a summary of collider signatures at the LHC for the $Z_2$ odd LQs and VLLs, in the context of existing LHC searches for such states.    

\section*{Acknowledgments}
SD and SKR would like to acknowledge the support from the Department of Atomic Energy (DAE), Government of India, for the Regional Centre for Accelerator based Particle Physics (RECAPP), Harish-Chandra Research Institute.
\bibliographystyle{JHEP}
\bibliography{Reference}%

\end{document}